\documentclass[11pt]{article}
\usepackage{amstext}
\usepackage{amsthm}
\usepackage{amsmath}
\usepackage{amssymb}
\usepackage{setspace}
\usepackage{enumerate}
\usepackage{graphicx}
\usepackage[toc,page]{appendix}
\usepackage{array}

\def\Proof{\par\noindent{\bf Proof:~}}
\def\blackslug{\hbox{\hskip 1pt \vrule width 4pt height 8pt
    depth 1.5pt \hskip 1pt}}
\def\QED{\quad\blackslug\lower 8.5pt\null\par}

\def\dnsitem{\vspace{-7pt}\item}
\def\dnssubitem{\vspace{-5pt}\item}

\newtheorem{theorem}{Theorem}[section]

\newtheorem{definition}[theorem]{Definition}

\newtheorem{claim}[theorem]{Claim}
\newtheorem{lemma}[theorem]{Lemma}

\newtheorem{corollary}[theorem]{Corollary}

\newtheorem{observation}[theorem]{Observation}

\newtheorem{property}[theorem]{Property}

\newtheorem{conjecture}{Conjecture}
\theoremstyle{definition}

\newtheorem{procedure}[theorem]{Procedure}

\def\propt1{\mathcal{P}_0}
\newtheorem*{tpropt1*}{Property $\propt1$}
\def\propbr{\mathcal{P}_1}
\def\propbbr{\mathcal{P}_2}
\def\formcto{\mathcal{F}_1}
\def\formctt{\mathcal{F}_2}
\def\regtwo{\mathcal{R}_1}
\def\regclean{\mathcal{R}_2}
\def\regct{\mathcal{R}_3}
\def\regcolored{\mathcal{R}_4}
\def\decomponereg{\mathcal{A}_1}
\def\decomponecorr{\mathcal{A}_2}
\def\decomproot{\mathcal{A}_3}
\def\decompsize{\mathcal{A}_4}
\def\decompcleanl{\mathcal{A}_5}
\def\decompedges{\mathcal{A}_6}
\def\boxpropnocorhigh{\mathcal{C}_1}
\def\boxpropnocorlow{\mathcal{C}_2}
\def\boxpropcor{\mathcal{C}_3}

\newcolumntype{C}[1]{>{\centering}m{#1}}

\def\invboxes{\mathcal{I}}
\newtheorem*{tinvboxes*}{Invariant $\invboxes$}
\def\invboxesdom{\mathcal{I}_D}
\def\invboxesstaller{\mathcal{I}_S}

\def\varexcess{\tt \Psi}

\def\performmatch{\mbox{\sf Perform\_Match}}
\def\play{\mbox{\sf Play}}
\def\update{\mbox{\sf Update}}
\def\densify{\mbox{\sf Densify}}
\def\interpart{\mbox{\sf InterPart}}
\def\joinhigh{\mbox{\sf JoinHigh}}
\def\discblue{\mbox{\sf DisconnectExtBlue}}
\def\fixbwparent{\mbox{\sf FixBWParent}}
\def\convhigh{\mbox{\sf ConvHigh}}

\begin{document}
\onehalfspacing

\title{The Domination Game: \\
Proving the 3/5 Conjecture on Isolate-Free Forests}
\date{\today}
\author{Neta Marcus
\thanks{The Weizmann Institute of Science, Rehovot, Israel.
{\tt \{neta.marcus,david.peleg\}@weizmann.ac.il}.}
\and
David Peleg $^*$ 
\thanks{Supported in part by the Israel Science Foundation (grant 1549/13)
and the I-CORE program of the Israel PBC and ISF (grant 4/11).}
}
\maketitle

\begin{abstract}
\pagenumbering{arabic} 
We analyze the \emph{domination game}, where two players, {\em Dominator} and {\em Staller}, construct together a dominating set $M$ in a given graph, by alternately selecting vertices into $M$. Each move must increase the size of the dominated set. The players have opposing goals: Dominator wishes $M$ to be as small as possible, and Staller has the opposite goal. 
Kinnersley, West and Zamani conjectured in \cite{kinnersley2013extremal} that when both players play optimally on an isolate-free forest, there is a guaranteed upper bound for the size of the dominating set that depends only on the size $n$ of the forest. This bound is $3n/5$ when the first player is Dominator, and $(3n+2)/5$ when the first player is Staller. 
The conjecture was proved for specific families of forests in \cite{kinnersley2013extremal} and extended by Bujt\'as in \cite{bujtas2015domination}. 
Here we prove it for all isolate-free forests, by supplying an algorithm for Dominator that guarantees the desired bound.
\end{abstract}

\section{Introduction}
\label{chapter:intro}

We analyze a two-party game on graphs called the \emph{domination game}, in which two players with opposing goals construct together a dominating set for a given graph.
The game was introduced by Bre\u{s}ar, Klav\u{z}ar and Rall in \cite{brevsar2010domination}.
One setting in which such a problem may be of interest is the following scenario:

{\em New city regulations state that a house is only fire-safe if it is a short distance from a trained firefighter.
In order to make sure all houses are fire-safe, a list of citizens that should be trained and hired as firefighters must be made.
Two people volunteer for the task of making the list:
The city treasurer, who wishes to minimize the costs and therefore wants the number of firefighters to be as small as possible,
and 
the head of the firefighters union, who benefits from adding new members and therefore wishes to maximize the number of firefighters.
The mayor, a seasoned politician, decides to let both volunteers add names to the list 
in turns, each adding a single firefighter that would improve the safety of the city by making additional houses fire-safe, until the new regulations are met.
What strategy should the treasurer adopt? What can be guaranteed about the outcome?
}

We study the possible outcomes of such a selection process under some specific settings.

\paragraph{The problem.}
Throughout, we consider an undirected graph $G(V,E)$ of size $|V|=n$.
We assume $G$ is \emph{isolate-free}, i.e., it has no isolated vertices.
A \emph{dominating set} is a set $S \subseteq V$ such that for each $v\in V$, either $v \in S$ or $v$ has a neighbor in $S$.

For a graph $G(V,E)$ and a subset $S \subseteq V$, denote the \emph{closed neighborhood} of $S$ by $\Gamma[S, G]$, that is,
$\Gamma[S, G] = S \cup \left\{ u \in V \mid  \mbox{there exists some } s \in S \mbox{ such that } (u, s) \in E \right\}$. 
For a single vertex $v \in V$, define 
$\Gamma[v, G] = \Gamma[\{v\}, G]$.
The \emph{open neighborhood} of $v$ is the set of its neighbors in $G$, and it is denoted by $\Gamma(v, G) = \left\{ u \in V \mid  (u, v) \in E \right\}$.
Whenever $G$ is clear from the context, we omit it and write simply $\Gamma[S]$, $\Gamma[v]$ or $\Gamma(v)$.
The size of the smallest dominating set in $G$ is denoted by $\gamma(G)$.
For any set $S$, a vertex $v \in \Gamma[S]$ is said to be \emph{dominated by $S$}.

In the \emph{domination game}, two players construct together a dominating set, $M$. 
The players alternate in taking turns, and in each turn, the current player picks a single vertex and adds it to $M$.
The two players are referred to as \emph{Dominator} and \emph{Staller}, and they have opposing goals regarding $M$ - Dominator wants to minimize $|M|$, while Staller wants to maximize it.

The chosen vertex at step $t$ is referred to as the player's \emph{move} in step $t$ or the $t$'th move, and is denoted by $m_t$.
The partial dominating set constructed at the end of step $t$ is denoted by $M_t = \{m_i : 1 \leq i \leq t\}$. Define $M_0 = \emptyset$.
A move $m_t$ is considered \emph{legal} if the dominated set increases, that is, $\Gamma[M_{t-1}] \subsetneqq \Gamma[M_t]$.
The players must make legal moves at all steps. 
The game ends when all vertices of $V$ are dominated by $M ( = M_T)$, that is, when $M$ is a dominating set of $G$.
Hence, the game is a maximal sequence of legal moves, that is, a sequence $(m_1, m_2, ..., m_T)$ such that $M_T$ is a dominating set but $M_{T-1}$ is not.

The domination game has two variants:
It is called a \emph{Dominator-start} game when the first move is taken by Dominator, 
and a \emph{Staller-start} game when Staller makes the first move.
Hence, in a Dominator-start game, the odd moves are decided by Dominator and the even moves are decided by Staller.
In a Staller-start game, it is the opposite.
When both players play optimally, we call the size of the resulting dominating set the \emph{game domination number} of $G$, 
and denote it by $\gamma_D(G)$ for the Dominator-start variant, and by $\gamma_S(G)$ for the Staller-start variant.

We wish to study the following conjecture, introduced in \cite{kinnersley2013extremal}.
\begin{conjecture}
\label{conjecture:3_5}
If $G$ is an isolate-free $n$-vertex forest (i.e., it has no singleton vertices), then 
$\gamma_D(G) \leq 3n/5$ 
and 
$\gamma_S(G) \leq (3n+2)/5$.
\end{conjecture}
The conjecture was later extended to general isolate-free graphs in \cite{bujtas2015domination}, but here we focus on forests.

Since our goal is to prove the conjecture, we introduce modified goals for the players.
We say Dominator wins in a Dominator-start (respectively, Staller-start) game if the game ends within at most $3n/5$ (resp., $(3n+2)/5$) moves, and otherwise Staller wins.

\begin{procedure}
\label{proc:game_outline}
Given an isolate-free $n$-vertex forest $G(V,E)$, the Dominator-start variant of the game can be described by the following algorithm.
\begin{enumerate}
	\dnsitem $M \leftarrow \emptyset$;
              $T_{\max} \leftarrow \left\lfloor 3n / 5 \right\rfloor$;
	      $t \leftarrow 0$
	\dnsitem While $\Gamma[M] \neq V$:
	\begin{enumerate}
		\dnsitem $t \leftarrow t + 1$
		\dnssubitem \emph{current player} $\leftarrow$ Dominator if $t$ is odd and Staller otherwise
		\dnssubitem Receive a legal move $v$ from the current player
		\dnssubitem $M \leftarrow M \cup \{v\}$.
	\end{enumerate}
	\dnsitem If $t \leq T_{\max}$, Dominator wins. Otherwise, Staller wins.
\end{enumerate}
\end{procedure}

Hereafter, the total number of moves in a specific execution of the game is denoted by $T$.

A similar algorithm can be used to describe the Staller-start variant, except that then $T_{\max}$ is set to $\left\lfloor (3n + 2) / 5 \right\rfloor$, and the odd moves are performed by Staller.

\paragraph{Previous approaches.}
As mentioned earlier, Conjecture \ref{conjecture:3_5} was introduced by Kinnersley, West and Zamani in \cite{kinnersley2013extremal}.
In the paper, the conjectured bound of $3 n / 5$ moves is achieved for specific types of forests, and a weaker bound of $7 n / 11$ moves is proved for arbitrary isolate-free $n$-vertex forests.

In \cite{bujtas2015domination}, Bujt\'as proves the conjecture for isolate-free forests in which no two leaves are at distance $4$, 
and improves the bound for arbitrary isolate-free $n$-vertex forests from $7 n / 11$ to $5 n / 8$.
The proofs in \cite{bujtas2015domination} use a method for coloring and evaluating vertices according to their state, and creating intermediate graphs, in order to choose moves and to prove the desired bound.

\begin{figure}[thbp]
  \caption{\sf Motivating example.}
  \medskip
  \centering
    \fbox{\includegraphics[width=7cm]{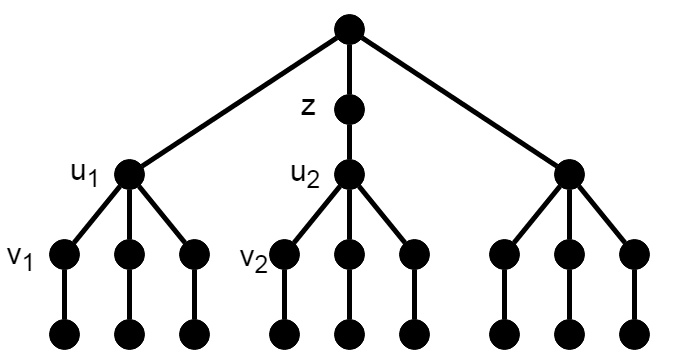}}
  \label{fig:locality_example_intro}
\end{figure}

\paragraph{Motivating example.}
We start with a simple example that illustrates some of the difficulties that any algorithm for Dominator must face.
Consider a Dominator-start game which is played on the graph shown in Figure \ref{fig:locality_example_intro}.
The graph contains $23$ vertices, therefore Dominator wins if and only if the game ends within $13$ moves or less.
Even though the neighborhoods of the vertices $v_1$ and $v_2$ are similar, 
the reader can verify that Dominator can win by playing $v_1$ or $u_1$ as the first move, 
whereas if Dominator plays $v_2$ or $u_2$, Staller can win the game by playing $z$ in the following move.

We believe this example can be extended to graphs of arbitrarily large size, in which choosing between moves that appear to be the same locally may determine the outcome of the game.

\paragraph{Our contributions.}
We provide an algorithm for Dominator that guarantees that the game ends within the number of moves required by Conjecture \ref{conjecture:3_5} on all isolate-free forests, which proves the conjecture.
We rely on the general method used in \cite{bujtas2015domination} and extend it, by separating the value of each vertex from its color, as well as fine-tuning additional aspects of the intermediate graphs. 
We start with Section \ref{section:dom_game_notations}, where we lay the foundations for the analysis by formalizing various aspects of the game.
We then introduce the algorithm in Section \ref{section:algorithm_outline}, and prove that it achieves the bound of Conjecture \ref{conjecture:3_5} in Section \ref{section:analysis}.
We conclude the analysis of Dominator's strategy in Section \ref{section:implementation}, where we discuss a possible implementation of the algorithm and describe our tests.
Finally, Section \ref{section:conclusions} contains some concluding remarks.

\section{Notation and Preliminaries}
\label{section:dom_game_notations}

Before describing the algorithm, we introduce some definitions and properties used to analyze the game.

\paragraph{Graph notions and vertex labeling.}
The graphs on which the game is played are undirected and unrooted forests that have no isolated (singleton) vertices.
We label all vertices in the initial graph $G$ with distinct even indices, $z_{2i}$.
The motivation for this is that we later introduce virtual vertices,
and we want an easy way to tell apart real (non-virtual) vertices from virtual ones - a real vertex $z_{2i}$ always has an even index, while a virtual vertex $z_{2j+1}$ has an odd index.

When we refer to \emph{components} of a graph, we always mean maximal connected components.
For a component $C$, define the size of $C$ to be the number of vertices in $C$ and denote it by $|C|$.
The degree of a vertex $v$ in a graph $G(V, E)$ is denoted by $d(v, G)$. When $G$ is clear from the context, it is denoted by $d(v)$.

\paragraph{Vertex color, vertex value and legal moves.}
Recall that the players construct together a set $M$, until it becomes a dominating set.
We use a variation of the grading system introduced in \cite{bujtas2015domination}.
During the game, each vertex has one of three possible \emph{colors}, and one of three possible \emph{values} 
(and these may change between steps).

\begin{definition}
Let $u$ be a vertex in the graph.
The \emph{color}, or \emph{type}, of $u$ at the end of step $t$, denoted by $c_t(u)$, is defined by the following properties:
\begin{itemize}
	\dnsitem $u$ is called \emph{white} or $c_t(u) = W$ if $u$ is not dominated, that is, $u \notin \Gamma[M_t]$.
	\dnsitem $u$ is called \emph{blue} or $c_t(u) = B$ if $u$ is dominated but has an undominated neighbor, that is, $u \in \Gamma[M_t]$ but $\Gamma[u] \not\subseteq \Gamma[M_t]$.
	\dnsitem $u$ is called \emph{red} or $c_t(u) = R$ if $u$ is dominated and all its neighbors are dominated, that is, $\Gamma[u] \subseteq \Gamma[M_t]$.
\end{itemize}
\end{definition}

When $t$ is clear from the context, we denote the color by $c(u)$ instead of $c_t(u)$.

We define $\mathcal{W}_t$, $\mathcal{B}_t$ and $\mathcal{R}_t$ to be the sets of vertices of type $W$, $B$ and $R$ (respectively) at the end of step $t$.

Even though the first step of the game is step $1$, we use (the end of) step $0$ to denote the state of the graph before the first move.

\begin{observation}
For all steps $t \geq 0$, $V = \mathcal{W}_t \cup \mathcal{B}_t \cup \mathcal{R}_t$, and these sets are disjoint.
\end{observation}

\begin{observation}
For every $t \geq 0$ and $v \in M_t$, $c_t(v) = R$.
Also, $\mathcal{W}_0 = V$ and $\mathcal{R}_T = V$, namely, $c_0(v) = W$ and $c_T(v) = R$ for every $v$.
\end{observation}

\begin{observation}
If $c_t(u) = B$ then $c_{t'}(u) \in \{B, R\}$ for all $t' > t$.
If $c_t(u) = R$ then $c_{t'}(u) = R$ for all $t' > t$.
\end{observation}

\begin{claim}
\label{claim:legal_moves}
At the beginning of step $t+1$, the set of legal moves consists of exactly the vertices of $\mathcal{W}_t \cup \mathcal{B}_t$, 
or in other words, the only vertices that a player cannot choose at step $t+1$ are those in $\mathcal{R}_t$.
This also implies that every move is either on or adjacent to some white vertex.
\end{claim}
\Proof
Let $v \in V$.
If $v$ is a red vertex, then $\Gamma[v] \subseteq \Gamma[M_t]$ and therefore $\Gamma[M_t \cup \{v\}] = \Gamma[M_t]$, so $v$ is an illegal move.
On the other hand,
if $v$ is a blue vertex, then $\Gamma[v] \not\subseteq \Gamma[M_t]$ and therefore ${\Gamma[M_t] \subsetneqq \Gamma[M_t \cup \{v\}]}$, so $v$ is a legal move.
Similarly, 
if $v$ is a white vertex, then $v \in \Gamma[M_t \cup \{v\}] \setminus \Gamma[M_t]$ and therefore $v$ is a legal move.
\QED

\begin{observation}
\label{obs:edges_no_r}
If for some step $t$, $c_t(u)=W$ and $c_t(v)=R$, then $(u,v) \notin E$ 
(that is, white and red vertices cannot be neighbors).
\end{observation}

\begin{definition}
For any step $t$ and for any $v \notin \mathcal{R}_t$, 
let $c_{t,v}(u)$ be the color of $u$ assuming the $(t+1)$st move was $v$, that is, $c_{t, v}(u) = c_{t+1}(u)$ if $m_{t+1} = v$.
\end{definition}

In addition to its color, each vertex also has a value.

\begin{definition}
A function $p : V \times \{0, ..., T\} \rightarrow \{0, 2, 3\}$ is called a \emph{value} function if it satisfies the following requirements for all $u \in V$.
\begin{itemize}
	\dnsitem If $c_t(u)=W$ then $p(u, t) = 3$,
	\dnsitem If $c_t(u)=R$ then $p(u, t) = 0$,
	\dnsitem If $c_t(u)=B$ then $p(u, t) \in \{2, 3\}$.
\end{itemize}
If $p(u, t)=k$, we say that at step $t$, $u$ is \emph{worth} $k$ points, or \emph{has value} $k$.
For a set of vertices $U \subseteq V$, define $p(U, t) = \sum_{u \in U}p(u, t)$. 
\end{definition}

When $t$ is clear from the context, we may omit it, and denote the value of $u$ by $p(u)$.

\begin{definition}
For any step $t$, a vertex $u$ is called \emph{high}, 
and its type is generically referred to as $H$, if $p(u, t) = 3$.
Let $\mathcal{H}_t$ denote the set of high vertices at the end of step $t$.
\end{definition}

\begin{definition}
For any step $t$ and for any vertex $u \in V$, if $c_t(u)=B$ and $p(u, t)=3$, we say that $u$ is a $B_3$ vertex (at the end of step $t$).
Similarly, if $c_t(u)=B$ and $p(u, t)=2$, then $u$ is called a $B_2$ vertex.
Note that saying that a vertex is of type $H$ is synonymous to saying that it is of type $W$ or $B_3$.
\end{definition}

In the graphical illustrations to appear hereafter, vertices are of type $H$ except where specifically labeled otherwise.

\begin{definition}
For any step $t$ and for any $v \notin \mathcal{R}_t$ and $u \in V$,
let $p(u, t, v)$ be the value of $u$ at the end of step $t+1$ assuming the $(t+1)$st move was $v$,
that is, $p(u, t, v) = p(u, t+1)$ if $m_{t+1} = v$.

\end{definition}

\begin{observation}
For any step $t$, for every $v \in M_t$ and for any value function $p(\cdot, \cdot)$, ${p(v, t) = 0}$.
Also, for any $u$ and for any value function, $p(u, 0) = 3$ and $p(u, T) = 0$, and consecutively ${p(V, 0) = 3n}$ and $p(V, T) = 0$.
\end{observation}

Let us remark that the value function defined later on for the algorithm will ensure that $p_t(V)$ is monotonically decreasing in $t$.

\paragraph{Gain.}
The \emph{gain} of a vertex $v$ under a given value function is the number of points gained when the current player chooses it. Formally, 
given the value function $p$, the corresponding gain function \\
${g : V \times \{1, ..., T\} \rightarrow \{0, ..., 3 \cdot |V|\}}$ is defined by
$g(v, t + 1) = p(V, t) - p(V, t, v)$.

Again, whenever $t$ is clear from the context, we omit it.

\begin{claim}
\label{claim:3pts}
For any $1 \leq t \leq T$ and for any $v \notin \mathcal{R}_{t-1}$, the following properties hold.
\begin{enumerate}[(a)]
{\setlength\itemindent{10pt} 
	\dnsitem 
	\label{claim:3pts:b}
	If $p(u, t) = 2$ for all $u \in \mathcal{B}_t \setminus \mathcal{B}_{t-1}$, then $g(v, t) \geq 3$.
	
	\dnsitem 
	\label{claim:3pts:w}
	If $c_{t-1}(v) = W$ then $g(v, t) \geq 3$ for any gain function $g(\cdot, \cdot)$.
}
\end{enumerate}
\end{claim}
\Proof
If $c_{t-1}(v)=W$, then for any value function it holds that $p(v, t-1) = 3$ and $p\left(v, t-1, v\right) = 0$. 
Therefore $g(v, t) \geq 3$ (for any gain function), establishing (\ref{claim:3pts:w}).
It remains to prove (\ref{claim:3pts:b}) in case $c_{t-1}(v)=B$. Then by the definition $v$ has some white neighbor, $u$, at the end of step $t-1$. 
Since $c_{t-1,v}(v) = R$, the value of $v$ decreases by at least $2$. 
Also, since $c_{t-1,v}(u) \in \{R,B\}$, we conclude that if $p$ assigns a value of $2$ to all new blue vertices then the value of $u$ decreases by at least $1$. 
Hence $g(v, t) \geq 2 + 1 = 3$.
\QED

\begin{corollary}
\label{cor:3pts_def}
It is always possible to define the value function such that at least $3$ points are gained in every legal move.
\end{corollary}
\Proof
Consider a move $m_t$. If $c_{t-1}(m_t) = W$, then $g(m_t, t) \geq 3$ since $c_t(m_t) = R$.
Otherwise, ${c_{t-1}(m_t) = B}$ and therefore $m_t$ has a neighbor $u$ such that $c_{t-1}(u) = W$.
The value of $m_t$ itself decreases by at least $2$. Hence if $c_t(u) = R$, then $g(m_t, t) \geq 2+3 = 5$ for any value function.
Otherwise $c_t(u) = B$, and we can choose $p(\cdot, \cdot)$ such that $p(u, t) = 2$, gaining $1$, and then ${g(m_t, t) \geq 2 + 1 = 3}$.
\QED
In fact, the algorithm will define the value function in such a way, namely, it will ensure that every move (including Staller moves) gains at least $3$ points.

We now formulate a useful condition on strategies. 
Denote the average gain per move (over the entire game) by
$$\hat{g} = \frac{1}{T} \cdot \sum_{t=1}^{T}g(m_t, t). $$
\smallskip
\par\noindent {\bf The average gain condition:}
The average gain per move satisfies $\hat{g}\ge 5$.
\begin{claim}
\label{claim:avg_5}
In a Dominator-start game, the average gain condition is equivalent to Conjecture \ref{conjecture:3_5}.
\end{claim}
\Proof
As $p(V, 0) = 3n$ and $p(V, T)=0$, we have $\hat{g} = \frac{3n}{T}$. 
Therefore, $T \leq \frac{3n}{5}$ is equivalent to 
$\frac{3n}{5} \cdot \hat{g} \geq T \cdot \hat{g} = 3n$,
 which yields $\hat{g} \geq 5$.
\QED

Next, denote the average gain over steps $2, ..., T$ by 
~~~~~ $\displaystyle \tilde{g} = \frac{1}{T-1} \cdot \sum_{t=2}^{T}g(m_t, t).$
\smallskip
\par\noindent {\bf The shifted average gain condition:}
Excluding the first move, the average gain satisfies $\tilde{g} \ge 5$.
\begin{claim}
\label{claim:staller_start_cond}
In a Staller-start game, the shifted average gain condition implies
Conjecture \ref{conjecture:3_5}.
\end{claim}
\Proof
By Claim \ref{claim:3pts}, $g(m_1, 1) \geq 3$, and if we use a value function $p$ satisfying $p(u, 1) = 3$ for all $u \in \mathcal{B}_1$, then 
$p(v, 1) = 3$ for all $v \notin \mathcal{R}_1$.
Assume $\tilde{g} \geq 5$.
Then $p(V, 1) \leq 3n - 3$ and $p(v, T) = 0$, and 
$$ g(m_1, 1) + (T-1)\cdot \tilde{g} = p(V, 0) - p(V, T) = 3n .$$
As $g(m_1, 1) \geq 3$ and $\tilde{g} \geq 5$, we have
$3 + (T-1) \cdot 5 \leq 3n$, or 
$T \leq \frac{3n-3}{5} + 1 = \frac{3n+2} {5}$,
establishing the claim.
\QED

\paragraph{Removing vertices and edges.}
Recall that red vertices are illegal moves and cannot be played, and are also worth $0$ points. Therefore we have the following.

\begin{observation}
Red vertices can be removed from the graph along with their edges, without changing the outcome of the game.
\end{observation}

By definition, each blue vertex $v$ has at least one white neighbor. 
Moreover, $v$ is converted from blue to red exactly when its last white neighbor is converted to blue or to red, regardless of the states of its blue neighbors. 
Therefore we have the following.
\begin{observation}
Edges between two blue vertices can be removed from the graph without changing the outcome of the game.
\end{observation}

However, it may sometimes be useful for our algorithm to keep edges that have a $B_3$ vertex as one of their endpoints.
The decision on whether to remove these edges or not will be made by the algorithm.

The algorithm maintains a graph called the \emph{underlying graph}, 
which contains only vertices and edges that may affect the outcome of the game.
This data structure also stores the decisions made by the algorithm about deleting edges between blue vertices,
and contains only the edges that were not deleted.
In particular, the algorithm ensures the following property, throughout the execution.

\begin{property}
\label{prop:underlying_graph}
The \emph{underlying graph} at the end of step $t$, denoted $G_t = (V_t, E_t)$, satisfies the following conditions.
\begin{enumerate}
	\dnsitem $G_0 = G$, and the vertices of $V$ are labeled with the labeling $z_{2i}$ defined in Section \ref{section:dom_game_notations}.
	\dnsitem $V_t$ = $\mathcal{W}_t \cup \mathcal{B}_t$.
	\dnsitem $E_t$ contains only edges that have at least one endpoint in $\mathcal{H}_t$ (this guarantees that both endpoints are in $V_t$ by Observation \ref{obs:edges_no_r}),
	and contains all edges that have at least one endpoint in $\mathcal{W}_t$.
	\dnsitem $E_t \subseteq E_{t-1}$.
\end{enumerate}
\end{property}

The following observation is an immediate result of the fact that edges are not removed as long as one of their endpoints is white.
\begin{observation}
\label{obs:w_neighborhood}
If $c_t(v) = W$, then $\Gamma[v, G_0] = \Gamma[v, G_t]$.
That is, the neighborhood of a white vertex does not change as long as it is  white (except maybe for some of its white neighbors turning blue).
\end{observation}

\begin{corollary}
\label{cor:last_5}
The last move on a component gains at least $5$ points.
\end{corollary}
\Proof
Since $G$ is isolate-free, and the last move is either on or adjacent to some white vertex (by Claim \ref{claim:legal_moves}), 
we conclude from Observation \ref{obs:w_neighborhood} that the underlying graph contains at least one additional vertex (that is not red) adjacent to the move. 
Therefore the total gain is at least $3+2=5$ points.
\QED

We want to define a single algorithm that will serve to prove the conjecture for both variants of the game.
The following corollary explains how this can be done.

\begin{corollary}
\label{cor:dom_stall_start}
Given an algorithm $A$, which guarantees that the game ends within at most $\frac{3n}{5}$ moves in the Dominator-start variant of the game given any initial isolate-free forest where all vertices are high (and not necessarily white), it is possible to construct an algorithm $B$ which guarantees that the game ends within at most $\frac{3n+2}{5}$ moves in the Staller-start variant of the game.
\end{corollary}
\Proof
The desired goal can be achieved by an algorithm $B$ that sets the value function at the end of the first step as described in the proof of Claim \ref{claim:staller_start_cond}, and then invokes $A$ for all the following moves (so that move $i$ is considered by $A$ as move $i-1$ for all $i \geq 1$).
This holds since the underlying graph $G_1$ contains only high vertices, so the corollary follows from Claim \ref{claim:staller_start_cond}.
\QED

Hereafter we focus on finding an algorithm which achieves the desired gain for the Dominator-start variant of the game, and the conjecture will follow from Corollary \ref{cor:dom_stall_start}.

\paragraph{Structural notations.}
We use the following definitions.

\begin{definition}[White, blue, high subgraph]
We say a subgraph of $G_t$ is white (respectively, blue, high) if all its vertices are white (respectively, blue, high).
Specifically, $G_0$ is high.
\end{definition}

\begin{definition}[Tail, Subtail] 
Let $P = (v_0, v_1, ..., v_k)$, $k \geq 1$, be a path in $G_t$,
that is, ${(v_i, v_{i+1}) \in E_t}$ for every $0 \leq i <k$.
$P' = (v_1, ..., v_k)$ is called a \emph{tail} if 
	$d(v_0) > 2, d(v_k) = 1$ (i.e., $v_k$ is a leaf), and $d(v_i) = 2$ for all $0<i<k$.
We call $v_1$ the \emph{tail lead}, and we say that $v_0$ has a tail.
If $d(v_0) \geq 1$, we say that $(v_1, ..., v_k)$ is a \emph{subtail}.
\end{definition}

\begin{figure}[thbp]
  \caption{\sf Graphical conventions used in our illustrations.}
  \medskip
  \centering
    \fbox{\includegraphics[width=9cm]{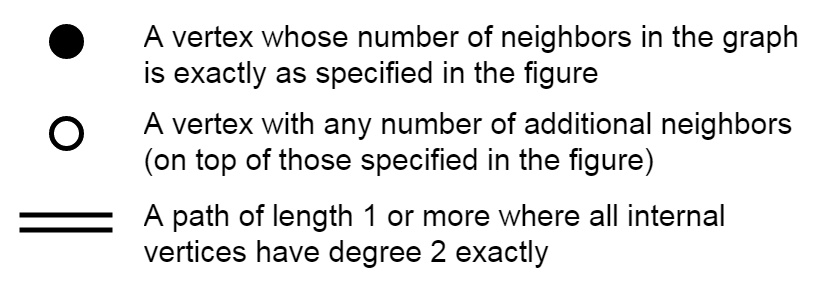}}
  \label{fig:notations}
\end{figure}

\begin{figure}[thbp]
  \caption{\sf Split vertex $v$ in general form.
Note that by our graphical conventions, the vertex $v$ has degree $3$ or higher, and $u$ has degree $1$ or higher, but $v_1$ and $v_2$ have degree exactly $1$.}
  \medskip
  \centering
    \fbox{\includegraphics[height=1.5cm]{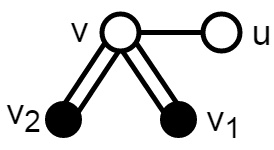}}
  \label{fig:splitP}
\end{figure}
	
\begin{definition}[Split vertex] 
\label{def:split_vertex}
A vertex of degree at least $3$ is called a \emph{split} vertex if it has at least two tails. See Figure \ref{fig:splitP} (our graphical conventions are summarized in Figure \ref{fig:notations}).
\end{definition}
	
\begin{definition}[Path component] 
A component is called a \emph{path component} if all its vertices have degree $1$ or $2$ 
(since the graph is a forest, there cannot be cycles).
Vertices on a path component that have degree $2$ are called \emph{internal vertices} of the component.

Path components may be described by a sequence of the colors of their vertices. 
For example, when we refer to ``\emph{a path of the form $B_2WH$}" we mean a path component of size $3$ with vertices $(v_1, v_2, v_3)$ such that $v_1$ is $B_2$, $v_2$ is white and $v_3$ is high.
Specifically, we use the term ``\emph{$BW$ component}" to describe a component of size $2$ containing one blue vertex and one white vertex. 
\end{definition}
	
\begin{definition}[Complex component] 
A component containing at least one split vertex is called a \emph{complex component}.
\end{definition}

\begin{claim}
\label{claim:every_tree_has_split}
Let $C$ be a component and let $r_1$ and $r_2$ be vertices in $C$ (not necessarily distinct).
If, when $C$ is rooted at $r_1$, the subtree $T$ rooted at $r_2$ is not a subtail, then $T$ contains a split vertex.
\end{claim}
\begin{figure}[thbp]
  \caption{\sf A sample subtree $T$ containing split vertices.}
  \medskip
  \centering
    \fbox{\includegraphics[width=7cm]{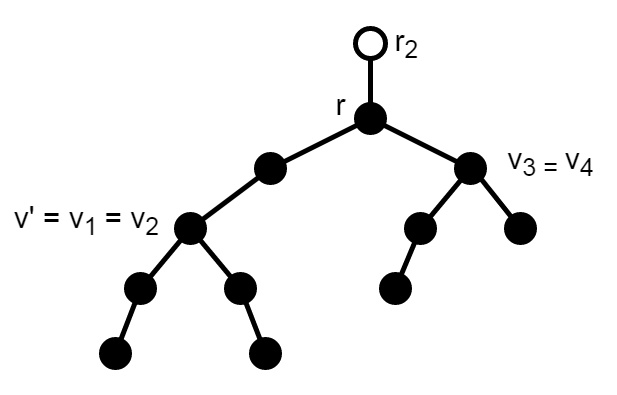}}
  \label{fig:subtree_splits}
\end{figure}
\Proof
Let $T$ be such a subtree. Let $r$ be a vertex on $T$ such that $d(r) \geq 3$ (guaranteed to exist since $T$ is not a subtail).
Let $\lambda_1, ..., \lambda_\ell$ be all the leaves of the subtree rooted at $r$, and for each $i$, let $v_i$ be the first vertex of degree at least $3$ on the (unique) path from $\lambda_i$ to $r$, including the endpoints (see an example in Figure \ref{fig:subtree_splits}). 
Since $d(r) \geq 3$, $v_i$ is guaranteed to exist for every $i$.
Notice that not all $v_i$ are distinct.
Let $v'$ be the $v_i$ farthest from $r$. 
Since $d(v') \geq 3$, there are at least two leaves, $\lambda_j$ and $\lambda_k$, in its subtree.
Since $v'$ is the first vertex on the path from $\lambda_j$ to $r$ that has degree at least $3$, we conclude that $v_j = v'$ and therefore $v'$ has a tail towards $\lambda_j$.
Similarly, we conclude that $v'$ has another tail towards $\lambda_k$.
Therefore $v'$ has at least two tails, which means it is a split vertex.
\QED

\begin{corollary}
Every tree containing a vertex of degree $3$ or more has at least one split vertex.
\end{corollary}

\begin{corollary}
Each (maximal connected) component is either a path component or a complex component.
\end{corollary}

\section{The algorithm}
\label{section:algorithm_outline}

\subsection{Outline}
\label{sub:alg_outline}

In order to prove Conjecture \ref{conjecture:3_5}, we show a possible course of action for every move that guarantees the average gain condition, namely, an average gain of $5$ points or more in the Dominator-start variant of the game.
By Corollary \ref{cor:dom_stall_start}, if Dominator uses an algorithm guaranteeing this gain, then Dominator wins both in the Dominator-start variant and in the Staller-start variant.

We do not describe a specific algorithm in this section, but rather show that such an algorithm exists.
In Section \ref{sub:algorithmic_details} we present a concrete naive algorithm resulting from this outline, and in Section \ref{section:implementation} we discuss better implementations.
Section \ref{sub:simplified_alg_no_distance_4} contains a simplified version of this algorithm, that can be used on isolate-free forests in which no two leaves are at distance $4$.

The suggested algorithm outline consists of several parts, performed for each move.
Suppose $t$ moves ($t < T$) were already played, and the algorithm needs to decide on the $(t+1)$st move (if it is a Dominator move), or preprocess for step $t+2$ (if $t+1$ is a Staller move).
\begin{enumerate}
	\dnsitem At the end of step $t$, the current underlying graph, denoted by $G_t$, undergoes a \emph{simulation} process consisting of two phases, each of which is described in detail later.
	\begin{itemize}
		\dnsitem {\bf Phase 1:}
		The graph is simplified by replacing subtrees of certain specific forms by virtual vertices (i.e., vertices that were not in $G_0$).
		The resulting (possibly smaller) graph is called the \emph{dense graph} and is denoted by $\hat{G}_t$.

		\dnssubitem {\bf Phase 2:}
		The resulting dense graph $\hat{G}_t$ is separated into \emph{boxes}, each of which is a connected subcomponent satisfying one of several properties.
		The process of separating the dense graph into boxes is called \emph{box decomposition}, and each vertex of the dense graph is assigned into a single box.
		We define Invariant $\invboxes$ which must be satisfied by the box decompositions used by the algorithm.
		A box decomposition satisfying this invariant is called a \emph{valid box decomposition}. 
		
		As becomes clear later, a dense graph may have more than one valid box decomposition, 
		and we show in the analysis that it is possible to maintain the underlying graph such that the corresponding dense graph has at least one valid box decomposition.
		We say that the underlying graph $G_t$ and the corresponding dense graph $\hat{G}_t$ are \emph{good} if $\hat{G}_t$ has a valid box decomposition, 
		and similarly we say that a component $C$ of the dense or underlying graph is \emph{good} if a graph containing only this component is good.
	\end{itemize}

	\dnsitem If move $m_{t+1}$ is performed by Staller, then the new underlying graph $G_{t+1}$ is generated from $G_t$ in a way that guarantees that at least $3$ points are gained 
	by Staller's move $m_{t+1}$, 
	and that the corresponding dense graph has a valid box decomposition.
	In the analysis, we show that an underlying graph satisfying these requirements can be generated from any good underlying graph and for any Staller move.

	\dnsitem 
	\label{alg_outline:choose_dom}
	Otherwise (move $m_{t+1}$ is a Dominator move),
	move $m_{t+1}$ is chosen (along with a corresponding underlying graph) greedily for Dominator from the vertices of $\hat{G}_t$,
	such that the gain is maximal among all such moves which result in a good underlying graph $G_{t+1}$.
	
	If several potential moves achieve the (same) maximal gain, ties are broken by choosing a move maximizing the minimal cumulative gain in the next three moves, 
	i.e., maximizing 
	$$
		\min_{m_{t+2}}{[g(m_{t+1}, t+1) + g(m_{t+2}, t+2) + g_{t+3}]} 
	$$
	where $g_{t+3}$ is the maximal gain that can be achieved by Dominator in its following move (with a good underlying graph), and we define $g_{t'} = 0$ for all $t' > T$.
	
	If there are still several such maximizing moves, then the tie is broken arbitrarily.
\end{enumerate}

It remains to describe the two phases of the simulation process.

\subsection{Phase $1$ of the simulation: Creating the dense graph}
\label{sub:dense_graph}

\begin{figure}[thbp]
  \caption{\sf All unlabeled vertices are white.
	      (a) A triplet subtree on the underlying graph. 
	      (b) The corresponding subtree on the dense graph. The triplet vertices in the set $\mathcal{TT}$ are the vertices that have numbers next to them. These numbers are their triplet depths.
	      The vertex $v$ is the only triplet head in (a), and its triplet witnesses are $v_1$, $v_2$ and $v_3$.
	      The vertices in $\mathcal{WT}_2$ in (a) are $v_1$, $v_2$ and all other vertices that are adjacent to leaves.
	      Note that we assume that $v$ does not have another neighbor in $\mathcal{TT}_3$ except for $v_1$ and $v_2$.
	     }
  \medskip
  \centering
    \fbox{\includegraphics[height=4cm]{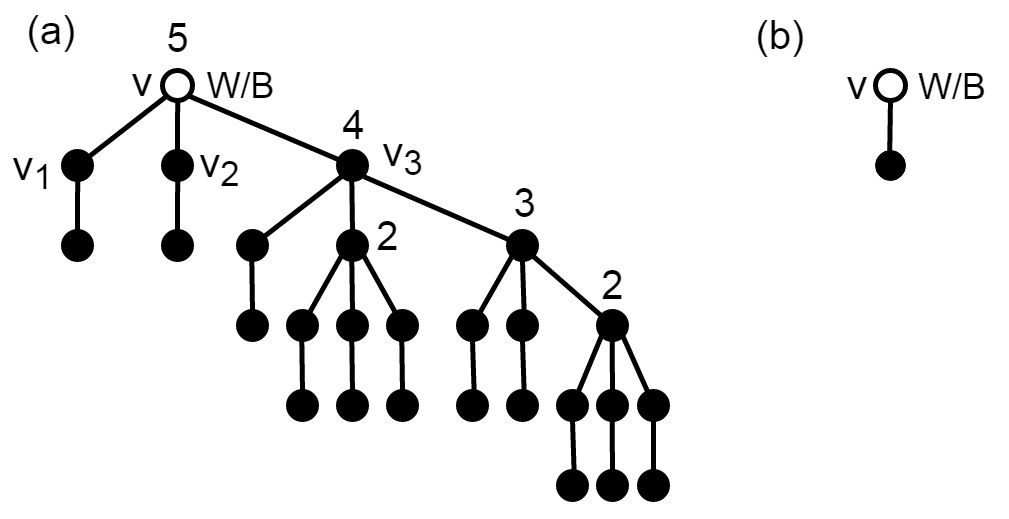}}
  \label{fig:t1_vertex}
\end{figure}

The dense graph is the result of removing subtrees called \emph{triplet witnesses} and replacing them with \emph{virtual leaves}.
The subtrees are constructed by the following process. 
Initially, set 
$$\mathcal{WT}_2 = \left\{ v \mid v \textrm{ is a lead of a white tail of length }2 \right\}.$$

Next,
the family $\mathcal{TT} = \bigcup_{i \geq 1} \mathcal{TT}_i$ of \emph{triplet vertices}, 
and the family $\mathcal{PW} = \bigcup_{i \geq 1} \mathcal{PW}_i$ of \emph{potential triplet witnesses},
are constructed using the following iterative rule.
For every $i \geq 1$, 
we construct in parallel the sets $\mathcal{TT}_i$ of triplet vertices and $\mathcal{PW}_i$ of potential triplet witnesses.
For each vertex $v \in \mathcal{TT}$ we also define its \emph{triplet depth}, $td(v)$, and its \emph{triplet subtree}.
We define $\mathcal{TT}_i$ and $\mathcal{PW}_i$ iteratively as follows.
\bigskip
\par\noindent
Initially,
$\mathcal{TT}_1 = \emptyset$; ~~
$\mathcal{PW}_1 = \mathcal{WT}_2$.

\bigskip
\par\noindent
After defining $\mathcal{TT}_i$ and $\mathcal{PW}_i$:
\begin{itemize}
	\dnsitem Add to $\mathcal{TT}_{i+1}$ every (blue or white) vertex $v$ that has at least three neighbors in $\mathcal{PW}_i$: \\
	${\mathcal{TT}_{i+1} = \mathcal{TT}_i \cup \left\{ v \in V \mid \left|\Gamma(v) \cap \mathcal{PW}_i \right| \geq 3 \right\}}$.
			
	\dnsitem Add to $\mathcal{PW}_{i+1}$ all vertices from $\mathcal{TT}_{i+1}$ that are white and have degree exactly $4$, and all vertices from $\mathcal{WT}_2$: \\
	${\mathcal{PW}_{i+1} = \left\{v \in \mathcal{TT}_{i+1} \mid c(v) = W \text{ and } d(v) = 4 \right\} \cup \mathcal{WT}_2}$.

	\dnsitem
	For each $v \in \mathcal{TT}_{i+1} \setminus \mathcal{TT}_i$ do:
	\begin{enumerate}
		\dnsitem Set the triplet depth of $v$, $td(v) = i + 1$.
		\dnssubitem Choose three \emph{triplet witnesses}, $v_1, v_2, v_3$, from the vertices of ${\Gamma(v) \cap \mathcal{PW}_i}$ (which is guaranteed to contain at least three vertices),
		according to the following priorities (in this order of significance):
		\begin{enumerate}
			\dnssubitem Prefer witnesses from $\mathcal{WT}_2$.
			\dnssubitem Prefer witnesses having lower triplet depth $td(v_i)$.
			\dnssubitem Prefer witnesses with higher $z_{2i}$ labels.
		\end{enumerate}

		The subtree containing $v$ and its witnesses (and no other neighbors of $v$) is called the \emph{triplet subtree rooted at $v$}.
	\end{enumerate}

\end{itemize}
Note that there is a maximal triplet depth $td_{\max} = \max_{v \in \mathcal{TT}} {td(v)}$ in the graph, and for all $i \geq td_{\max}$, ${\mathcal{TT}_i = \mathcal{TT}_{td_{\max}}}$ (while $\mathcal{TT}_{td_{\max} - 1} \subsetneqq \mathcal{TT}_{td_{\max}}$).
This is true since the graph's diameter upper bounds $td(v)$ for every $v$.
If $\mathcal{TT} = \emptyset$, then set $td_{\max} = 0$.

See illustration in Figure \ref{fig:t1_vertex}.

\begin{definition}
Let $v \in V$ be a triplet vertex.
If $v$ is not a triplet witness, then it is called a \emph{triplet head}
(note that $v$ may still be a potential triplet witness that was not chosen as a witness).
\end{definition}

\begin{observation}
\label{obs:triplet_subtrees_white}
Let $v \in V$ be a triplet vertex. 
All vertices in the triplet subtree rooted at $v$ are white, except (possibly) for $v$ itself, which is either white or blue.
If $v$ is not a triplet head, then $v$ is white as well.
\end{observation}

\begin{claim}
\label{claim:triplet_vertex_implies_head}
Let $C$ be a tree. If $C$ contains a triplet vertex, then it contains a triplet head.
\end{claim}
\Proof
Consider the set $\mathcal{TT}_C = \mathcal{TT} \cap C$, and let $v$ be a vertex in $\mathcal{TT}_C$ with maximal triplet depth (among the vertices of $\mathcal{TT}_C$).
By the way we define $td(v)$ we know that $v \in \mathcal{TT}_{td(v)} \setminus \mathcal{TT}_{td(v) - 1}$, 
and since $\mathcal{PW}_i \subseteq \mathcal{TT}_i \cup \mathcal{WT}_2$ for all $i$, we conclude that $v$ is not a triplet witness, and therefore it is a triplet head.
\QED

\begin{definition}
A \emph{virtual vertex} or \emph{virtual leaf} is a white leaf with odd label $z_{2i+1}$ that exists only on the dense graph, and is adjacent to a vertex $z_{2i}$ that is a triplet head on the underlying graph.
A vertex that is not virtual is called \emph{real}, and each real vertex has at most one virtual neighbor.
\end{definition}

The dense graph is created by replacing all triplet witnesses of each triplet head, along with their entire subtrees, 
with a single virtual vertex colored white (see Figure \ref{fig:t1_vertex}(b)), thus converting each triplet subtree into a subtail of length $2$.
This operation can be performed as follows:
\smallskip
\par\noindent{\bf Procedure \densify}:
\begin{enumerate}
	\dnsitem
	Calculate $\mathcal{T} = \left\{z \in G_t \mid z \textrm{ is a triplet head}\right\}$.
	
	/* \textit{Note that $G_t$ does not contain virtual vertices.} */ 
	\dnsitem For each $z_{2i} \in \mathcal{T}$:
	\begin{enumerate}
		\dnsitem Disconnect the edges between $z_{2i}$ and its triplet witnesses, and remove the components containing the triplet witnesses.
		\dnssubitem Create a new (virtual) white leaf $z_{2i+1}$ and add an edge between $z_{2i}$ and $z_{2i+1}$.
	\end{enumerate}
	\dnsitem Return the resulting graph.
\end{enumerate}
The dense graph $\hat{G}_t$ results from invoking the procedure $\densify$ on $G_t$.

\subsection{Phase $2$ of the simulation: Box decomposition}
\label{sub:alg_decomp}

In the second phase of the simulation, the algorithm decomposes the dense graph $\hat{G}_t$ into boxes, so that each vertex belongs to exactly one box.
We start by defining the boxes and their possible types.

\subsubsection{Box types}
\label{sub:box_types}

We now define a box, and the four possible box types.

\begin{definition}
\label{def:box_general}
Let $\hat{V}_t$ be the set of vertices in $\hat{G}_t$, and let $Q \subseteq \hat{V}_t$ be a connected subset of vertices in the dense graph.
$Q$ is a \emph{box} in $\hat{G}_t$ if it satisfies the following requirements.
\begin{enumerate}
	\dnsitem $Q$ is of (at least) one of four types: \emph{regular}, \emph{dispensible}, \emph{high leftover} and \emph{corrupted}, which are defined below.
	\dnsitem $Q$ contains at most two $B_2$ vertices.
	\dnsitem If $Q$ is not regular, then it has a blue vertex $r$ called the \emph{box root}, and $r$ does not have a neighbor in $Q$ that is a (white) leaf. 
\end{enumerate}
For a vertex $v \in Q$, we define the \emph{internal neighbors} of $v$ to be its neighbors inside the box, and the \emph{internal degree} of $v$ to be the number of internal neighbors it has.
\end{definition}

From now on,
whenever we consider the degree or the neighbors of a vertex in a specific box, 
we mean its internal degree and its internal neighbors, 
except where specifically noted otherwise.

\begin{figure}[thbp]
  \caption{\sf (a) Dispensible box of type $1$. 
	      (b) Dispensible box of type $2$ where condition \ref{def:dispensible:d2:p1} holds.
	      (c) Dispensible box of type $2$ where condition \ref{def:dispensible:d2:p2} holds.
	      Box roots are marked as $r$.}
  \medskip
  \centering
    \fbox{\includegraphics[height=4cm]{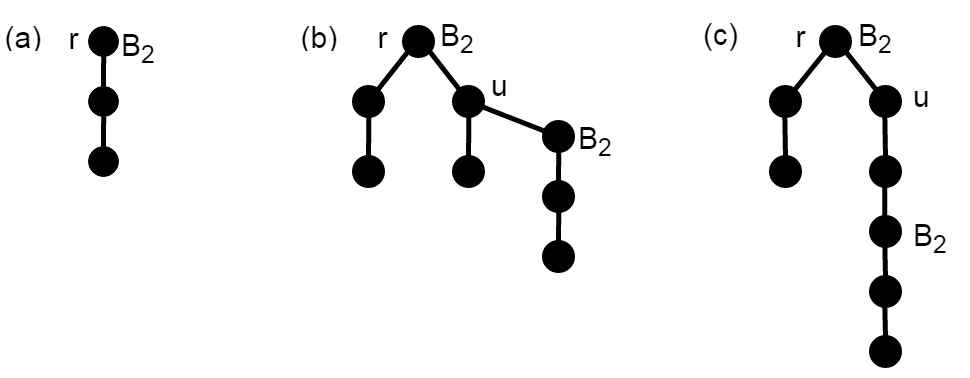}}
  \label{fig:dispensible}
\end{figure}

\begin{definition}
\label{def:dispensible}

There are two types of dispensible boxes.
\begin{enumerate}
	\dnsitem A box $Q$ is called \emph{dispensible of type $1$}, denoted by $D_1$, if it is a path $(v_1, v_2, v_3)$ of the form $B_2HH$, and the box root is $v_1$.

	\dnsitem A box $Q$ of size $8$ is called \emph{dispensible of type $2$}, denoted by $D_2$, if the following conditions hold.
	\begin{enumerate}
		\dnsitem The box root $r$ is a $B_2$ vertex of internal degree $2$.
		\dnssubitem $r$ has a high subtail of length $2$.
		\dnssubitem The neighbor $u$ of $r$ that is not on the high subtail satisfies exactly one of the following conditions.
		\begin{enumerate}
			\dnsitem 
			\label{def:dispensible:d2:p1}
			$u$ has internal degree $3$, and it has two additional neighbors in $Q$, $\lambda$ and $u'$, such that $\lambda$ is a high leaf, 
			and $u'$ is the ($B_2$) lead of a tail of the form $B_2HH$ (note that this implies that $u'$ could be the root of a $D_1$ box).
			\dnssubitem 
			\label{def:dispensible:d2:p2}
			$u$ has internal degree $2$, and it is the lead of a subtail of the form $HHB_2HH$ (in this case as well, the $B_2$ vertex on the tail could be the root of a $D_1$ box).
		\end{enumerate}
	\end{enumerate}
\end{enumerate}
 
A box is called \emph{dispensible}, denoted by $D$, if it is dispensible of type $1$ or $2$.
See Figure \ref{fig:dispensible} for illustrations.
\end{definition}

\begin{definition}
A box $Q$ in $\hat{G}_t$ is called a \emph{high leftover box} if all its vertices are high and it has a $B$ root,
and additionally, it does not contain triplet subtrees. 
\end{definition}

\paragraph{}
There are several types of regular boxes, defined below.

\begin{figure}[thbp]
  \caption{\sf Boxes corresponding to the different types of regular colored boxes (not all requirements are illustrated).
		(a) A box satisfying Property $\propbr$:(\ref{invnp:1leaf}).
		(b) A box satisfying Property $\propbr$:(\ref{invnp:1tail}).
		(c) A box satisfying Property $\propbr$:(\ref{invnp:d1}).
		(d) A box satisfying Property $\propbbr$:(\ref{invnp:2leaves}).
		(e) A box satisfying Property $\propbbr$:(\ref{invnp:2tail}).}
  \medskip
  \centering
    \fbox{\includegraphics[width=14cm]{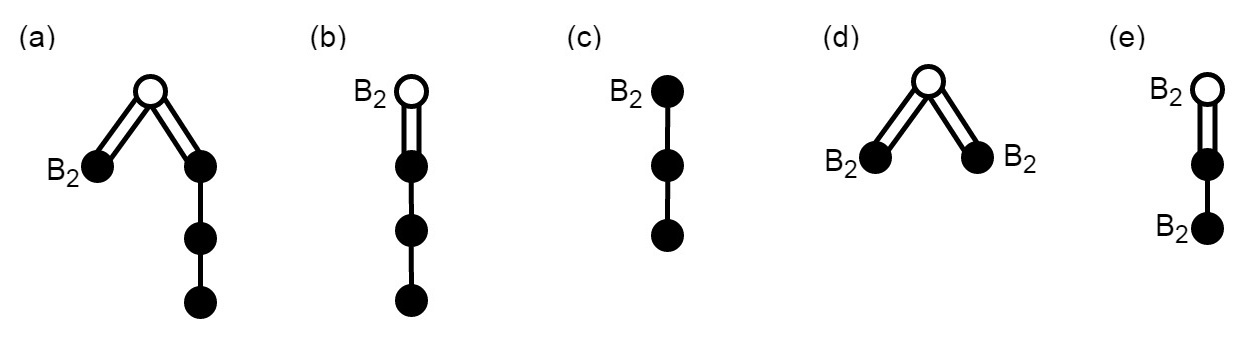}}
  \label{fig:regular_complex}
\end{figure}

\begin{definition}
\label{def:regular_colored}

A box $Q$ of size $3$ or more is called a \emph{regular colored} box if it satisfies exactly one of the following two properties, $\propbr$ and $\propbbr$, and additionally it satisfies Property $\propt1$ defined below.

\begin{description}
	\dnsitem[$\propbr$.] $Q$ contains a single $B_2$ vertex, $v$, such that at least one of the following conditions is satisfied. 
	\begin{enumerate}[(a)]
		\dnsitem \label{invnp:1leaf}
		$v$ is a leaf on a subtail of a vertex $u$, and $u$ has a high subtail of length $3$ or more and does not have white subtails of length $1$ or $2$. 
		\dnssubitem \label{invnp:1tail}
		$v$ has a (high) subtail of length $3$ or more, and no leaf neighbors. 
		\dnssubitem \label{invnp:d1}
		$v$ is a leaf and $|Q| = 3$ (i.e., $Q$ is a dispensible box of type $1$).
	\end{enumerate}

	\dnsitem[$\propbbr$.] $Q$ contains two $B_2$ vertices, $v_1$ and $v_2$, 
	such that the internal degree of $v_1$ is not greater than the internal degree of $v_2$,
	and at least one of the following conditions holds. 
	\begin{enumerate}[(a)]
		\dnsitem \label{invnp:2leaves}
		$v_1$ and $v_2$ are leaves of subtails of the same vertex $u$, and $u$ does not have a white leaf.
		\dnssubitem \label{invnp:2tail}
		$v_1$ is a leaf of a subtail of $v_2$, and $v_2$ does not have leaf neighbors.
	\end{enumerate}
\end{description}

See Figure \ref{fig:regular_complex} for illustrations, and note that regular colored boxes do not necessarily contain a split vertex. 

\begin{tpropt1*}
Let $v$ be a triplet vertex of depth $2$ in a box $Q$ of the dense graph. 
Then for every three white tails of length $2$ of $v$ whose tail leads 
are not all in $\mathcal{PW}$ (i.e., not all three tail leads are potential triplet witnesses in the underlying graph $G_t$), 
at least one vertex $v' \neq v$ in one of these tails is the parent of a box $Q'$ whose box root has internal degree at most $1$
(i.e., $Q'$ is either a dispensible box of type $1$, or a high leftover or corrupted box whose box root has at most one internal neighbor).
Note that this relates to all white tails of length $2$ of $v$, and not only the tails lead by the current triplet witnesses.
\end{tpropt1*}
\end{definition}

\begin{figure}[thbp]
  \caption{\sf (a) $C_{12}$ box of Form $\formcto$. 
	      (b) $C_{12}$ box of Form $\formctt$. 
	}
  \medskip
  \centering
    \fbox{\includegraphics[height=4cm]{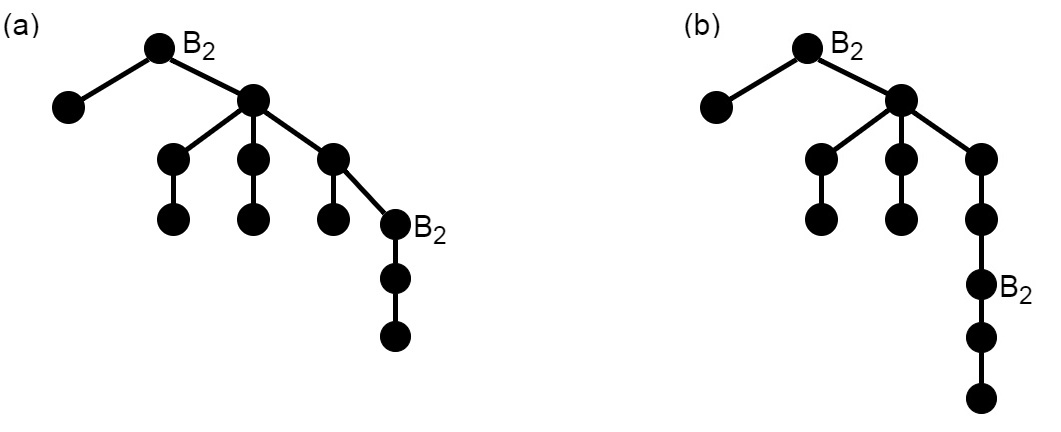}}
  \label{fig:c12_box}
\end{figure}

\begin{definition}
A box $Q$ is called a \emph{$C_{12}$ box} if it contains exactly $12$ vertices and is of one of the forms $\formcto$ or $\formctt$ (see Figure \ref{fig:c12_box}):
\begin{description}
	\dnsitem[$\formcto$.]
	$Q$ contains two high split vertices, $v_1$ and $v_2$, which are neighbors, and have the following tails:
	\begin{enumerate}[(a)]
		\dnsitem $v_1$ has a $B_2W$ tail and two high tails of length $2$.
		\dnssubitem $v_2$ has a high leaf, and a $B_2HH$ tail.
	\end{enumerate}

	\dnsitem[$\formctt$.]
	$Q$ contains a single high split vertex with exactly four tails of the following forms:
	\begin{enumerate}[(a)]
		\dnsitem Two high tails of length two.
		\dnssubitem A $B_2W$ tail.
		\dnssubitem A tail of length $5$ of the form $HHB_2HH$.

	\end{enumerate}
\end{description}
\end{definition}

\begin{definition}
\label{def:regular_box}
A \emph{regular box} is a box $Q$ that does not have a box root and satisfies exactly one of the following properties.
\begin{description}
	\dnsitem[$\regtwo$.] $Q$ is of size $2$.
	\dnsitem[$\regclean$.] $Q$ is high, i.e., does not contain $B_2$ vertices, and it contains at least $3$ vertices and satisfies Property $\propt1$.
	\dnsitem[$\regct$.] $Q$ is a $C_{12}$ box.
	\dnsitem[$\regcolored$.] $Q$ is a regular colored box.
\end{description}

If $Q$ contains a split vertex and is not a $C_{12}$ box, it is called a \emph{regular complex} box.
If $Q$ does not contain split vertices, it is called a \emph{regular path} box.
\end{definition}

\begin{definition}
A box $Q$ that is not dispensible, high leftover or regular is called a \emph{corrupted box}. 
\end{definition}

\subsubsection{The decomposition}

\begin{figure}[thbp]
  \caption{\sf A component of the dense graph and a valid box decomposition. Boxes are marked by dotted rectangles.}
  \medskip
  \centering
    \fbox{\includegraphics[height=8cm]{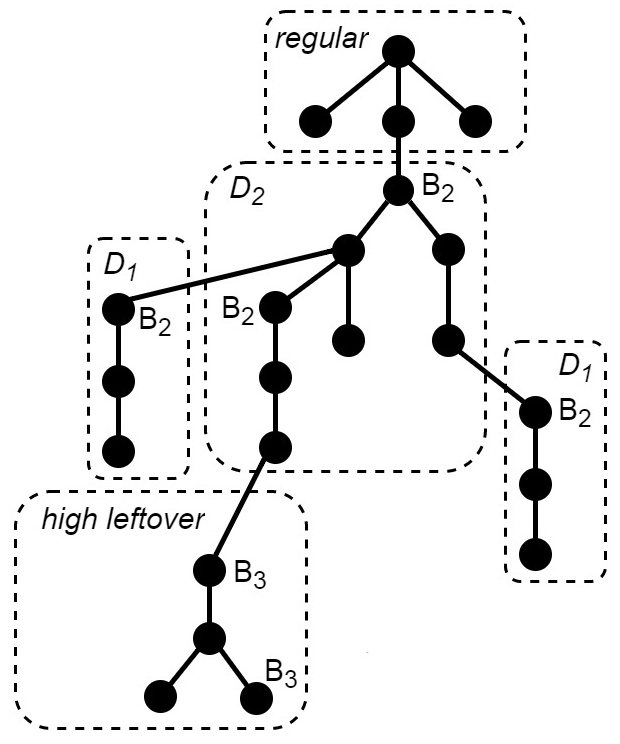}}
  \label{fig:box_decomp}
\end{figure}

In the second phase of the simulation, each maximal connected component $C$ in the dense graph $\hat{G}_t$ is decomposed into boxes according to the following definition.
\begin{definition}
\label{def:box_decomp}
Let $\hat{V}_t$ be the set of vertices in $\hat{G}_t$, and let $\mathcal{Q} = \left\{Q_1, Q_2, ...\right\}$ be a partition of $\hat{V}_t$ into boxes, i.e., a collection of subsets of $\hat{V}_t$ satisfying Definition \ref{def:box_general}, such that:
\begin{enumerate}[(a)]
	{\setlength\itemindent{10pt} \dnsitem $\bigcup_{Q_i \in \mathcal{Q}}{Q_i} = \hat{V}_t$.}
	{\setlength\itemindent{10pt} \dnsitem $Q_i \cap Q_j = \emptyset$ for every $Q_i, Q_j \in \mathcal{Q}$, $Q_i \neq Q_j$.}
\end{enumerate}
$\mathcal{Q}$ is called a \emph{box decomposition} of $\hat{G}_t$ if it satisfies the following properties.
\begin{description}
	\dnsitem[$\decomponereg$.] Each connected component of $\hat{G}_t$ contains at most one regular box.
	\dnsitem[$\decomponecorr$.] $\mathcal{Q}$ contains at most one corrupted box.
	\dnsitem[$\decomproot$.] For each box $Q_i$ that is not regular, the box root $r$ of $Q_i$ has at most one neighbor outside the box, and if such a neighbor $p$ exists, then it is not a box root.
	The vertex $p$ (if exists) is called $r$'s \emph{parent}, and the box $P$ containing $p$ is called the \emph{parent box} of $Q_i$. 
	The single box in each component that does not have a parent is called the \emph{root box} of the component.
	\dnsitem[$\decompsize$.] All parent boxes are of size $3$ or more. 
	\dnsitem[$\decompcleanl$.] The parent box of a high leftover box is not high.
	\dnsitem[$\decompedges$.] Edges between boxes always connect a box root to its parent, and are called \emph{external edges}.  
\end{description}

\end{definition}

Component types are defined according to their root boxes. 
For example, a component whose root box is dispensible is called a dispensible component. 
One exception is that a component is called a \emph{corrupted component} if it contains a corrupted box anywhere in it (regardless of the type of its root box).

\paragraph{}
Finally, we define a semi-corrupted component.

\begin{definition}
A component $C$ of the dense graph $\hat{G}_t$ is called \emph{semi-corrupted} if every one of its box decompositions contains a corrupted box, 
but there exists some $v \in C$ such that if $m_{t+1} = v$, then there exist an underlying graph $G_{t+1}$, a corresponding dense graph $\hat{G}_{t+1}$ and a box decomposition $\mathcal{Q}$ of $\hat{G}_{t+1}$ that does not contain corrupted boxes, and the gain from playing $v$ on $G_t$ is at least $8$ points.
\end{definition}

We are now ready to introduce the invariant that must be maintained by the algorithm.

\begin{tinvboxes*}
Let $t$ be any step in the game.
\begin{description}
	\dnsitem[$\invboxesdom$.] If move $m_t$ is played by Dominator, then there exists a box decomposition $\mathcal{Q}$ of the dense graph $\hat{G}_t$ that does not contain corrupted boxes.
	\dnsitem[$\invboxesstaller$.] If move $m_t$ is played by Staller, then there exists a box decomposition $\mathcal{Q}$ of the dense graph $\hat{G}_t$ that contains at most one corrupted box, 
	and if such a box exists then it is in a semi-corrupted component.
\end{description}
\end{tinvboxes*}

\begin{definition}
\label{def:good_graph_valid_box_decomp}
$ $
\begin{enumerate}
	\dnsitem The underlying graph $G_t$ and the corresponding dense graph $\hat{G}_t$ are \emph{good} if $\hat{G}_t$ has a \emph{valid} box decomposition, i.e., a box decomposition satisfying Invariant $\invboxes$.
	\dnsitem A component $C$ of the dense graph $\hat{G}_t$ or the underlying graph $G_t$ is \emph{good} if a graph containing only this component is good.
\end{enumerate}
\end{definition}

See Figure \ref{fig:box_decomp} for an example component on the dense graph and a valid box decomposition.

In Section \ref{section:analysis} we show that there exists an algorithm following the described outline, such that for every $t$, if at the end of step $t$ the dense graph is good (i.e., it satisfies Invariant $\invboxes$), and the average gain up to (and including) step $t$ is at least $5$ points, then the algorithm guarantees that the average gain at the end of 
some future step $t' > t$ 
is at least $5$ points.
This, in turn, guarantees also that the average gain at the end of the game is at least $5$ points.

\subsection{Algorithmic details}
\label{sub:algorithmic_details}
The outline described in the previous subsections gives rise to the following naive (and highly inefficient) implementation.

\smallskip
\par\noindent{\bf Procedure $\performmatch(G^*(V^*, E^*))$:}
\begin{enumerate}
	\dnsitem Initialize the state:
	\begin{enumerate}
		\dnsitem $G \leftarrow G^*$ with a fixed vertex labeling (see Section \ref{section:dom_game_notations}.
/*  \textit{the underlying graph} */
		\dnssubitem $M \leftarrow \emptyset$. /* \textit{the constructed dominating set} */
		\dnssubitem $p(v) \leftarrow 3$ for every $v \in V$. /* \textit{the value function} */
		\dnssubitem $t \leftarrow 0$. /* \textit{counter of moves} */
	\end{enumerate}
	
	\dnsitem While $V \neq \emptyset$:
	\begin{enumerate}
		\dnsitem $t \leftarrow t + 1$.
		
		\dnssubitem If $t$ is odd: /* \textit{Dominator's turn} */
		\begin{enumerate}
			\dnsitem Create the dense graph $\hat{G}$ from $G$ using Procedure $\densify$ of Section \ref{sub:dense_graph}.
			\dnsitem $v \leftarrow \play(\hat{G})$. /* \textit{choose a move for Dominator and update the underlying graph and value function} */ 
		\end{enumerate}
		
		\dnssubitem Else: /* \textit{Staller's turn} */
		\begin{enumerate}
			\dnsitem Receive a legal move $v$ from Staller.
			\dnsitem $\update(v)$. /* \textit{update the underlying graph and value function} */
		\end{enumerate}
		\dnssubitem $M \leftarrow M \cup \{v\}$.
	\end{enumerate}
\end{enumerate}

The procedure $\update(v)$ performs the following actions:
\begin{enumerate}
	\dnsitem For each choice of a pair $(G', p')$ of an underlying graph $G'$ satisfying Property \ref{prop:underlying_graph} and a value assignment $p'$, 
	which may result from the move $v$,
	check if $G'$ with the value assignment $p'$ has a valid box decomposition.
	\dnsitem From the collection of pairs $(G', p')$ which have a valid box decomposition,
	choose the pair $(\widetilde{G}, \tilde{p})$ achieving the highest excess gain, i.e., maximizing $p(G) - \tilde{p}(\widetilde{G})$ (breaking ties arbitrarily),
	and update $G$ and $p$ accordingly.
\end{enumerate}

The procedure $\play(\hat{G})$ performs the following actions:
\begin{enumerate}
	\dnsitem $C \leftarrow \emptyset$.
	\dnsitem For each vertex $m \in \hat{G}$:
	\begin{itemize}
		\dnsitem[] For each choice of a pair $(G_m, p_m)$ of an underlying graph $G_m$ satisfying Property \ref{prop:underlying_graph} and a value assignment $p_m$, 
		which may result from the move $m$:
		\begin{itemize}
			\dnssubitem[] If $G_m$ with $p_m$ has a valid box decomposition,
			then add $(m, G_m, p_m)$ to $C$.
		\end{itemize}
	\end{itemize}
	\dnsitem From the collection $C$, 
	choose the sequence $(\widetilde{m}, \widetilde{G}, \tilde{p})$ achieving the highest gain, i.e., maximizing $p(G) - \tilde{p}(\widetilde{G})$, breaking ties as described in Step \ref{alg_outline:choose_dom} in Section \ref{sub:alg_outline} (by checking all possible choices for the next three moves and choosing the move which maximizes the minimal gain over these moves).
	\dnsitem Update $G$ and $p$ according to $\widetilde{G}$ and $\tilde{p}$.
	\dnsitem Return $\widetilde{m}$ as the selected move for step $t$.
\end{enumerate}

\paragraph{}
Section \ref{section:implementation} contains a short discussion of more efficient implementations for the strategy outlined in Sections \ref{sub:alg_outline} through \ref{sub:alg_decomp}.

\subsection{Simplified algorithm for forests in which no two leaves are at distance $4$}
\label{sub:simplified_alg_no_distance_4}

Conjecture \ref{conjecture:3_5} was proved in \cite{bujtas2015domination} for isolate-free forests in which no two leaves are at distance exactly $4$, using a much simpler algorithm than the one described in this thesis.
We note that our algorithm can also take a simpler form when used on this family of graphs.
It may be instructive to consider this variant, in order to pinpoint the aspects of our algorithm that were needed in order to handle the possible existence of pairs of leaves at distance 4.

Observe that if no two leaves are at distance $4$ from each other, then $G$ does not contain triplet subtrees.
Therefore, there is no difference between the underlying graph $G_t$ and the dense graph $\hat{G}_t$ (and since Property $\propt1$ refers to triplet subtrees, this property also becomes irrelevant).
Additionally, it is always possible for Dominator to make a move gaining at least $7$ points on any regular colored or high box whose size is greater than $2$ (this fact follows from Claims \ref{claim:b2_dom} and \ref{claim:dom_split_7} which appear later in the analysis of Dominator moves), and any Staller move on such boxes gains at least $3$ points (by Claim \ref{claim:staller_regular} in the analysis of Staller moves),
and in both cases all resulting boxes can be regular boxes that are not $C_{12}$ boxes.
As a result, there is no need to perform the box decomposition, and the following simpler algorithm suffices (compare this to the algorithm in Section \ref{sub:alg_outline}):

Suppose $t$ moves ($t < T$) were already played, and the algorithm needs to decide on the $(t+1)$st move (if it is a Dominator move), or preprocess for step $t+2$ (if $t+1$ is a Staller move).
\begin{enumerate}
	\dnsitem If move $m_{t+1}$ is performed by Staller, then the new graph $G_{t+1}$ is generated from $G_t$ in a way that guarantees that at least $3$ points are gained 
	by Staller's move $m_{t+1}$, 
	and that all components of $G_{t+1}$ are regular complex or regular path boxes.

	\dnsitem 
	Otherwise (move $m_{t+1}$ is a Dominator move),
	move $m_{t+1}$ is chosen greedily for Dominator from the vertices of $G_t$,
	such that the gain is maximal among all such moves which result in a graph $G_{t+1}$ all of whose components are regular complex or regular path boxes. 
	If there are ties, they are broken arbitrarily.	
\end{enumerate}
Note that when all components are of size $2$, all moves (by both players) gain at least $5$ points.

\paragraph{}
The complex form of the algorithm for general isolate-free forests results from the fact that in some cases, Dominator must play on graphs that only contain split vertices with two or three tails, and all these tails are white tails of length $2$. 
This prompted the creation of the dense graph (for handling triplet subtrees), as well as the addition of dispensible boxes, which contain subtrees that can be ignored by Dominator (i.e., that cannot reduce the gain of playing on their parent boxes. See Lemma \ref{claim:ana_box} for details). 
High leftover and corrupted boxes were added in order to handle Staller moves on these ``hidden" subtrees, while Property $\propt1$ and $C_{12}$ boxes were added in order to make sure Dominator can always make moves that achieve the desired gain.
The analysis section that follows covers all possible moves on all types of boxes, but the core of the algorithm remains this simplified version.

\section{Analysis}
\label{section:analysis}

We separate the analysis into several parts, and show that Invariant $\invboxes$ is always satisfied, and that an average gain of $5$ points per move is achieved.
In Section \ref{sub:an_gain} we introduce sufficient conditions for guaranteeing Dominator's win.
In Section \ref{sub:an_simplify}  we show properties that simplify the case analysis,
and Section \ref{sub:special_subtrees} contains an analysis of two special subtrees that appear in many of the cases. 
We then analyze all possible moves and the resulting graphs.
The possible outcomes of Staller moves are analyzed in Section \ref{sub:an_staller}, 
and those of Dominator moves in Section \ref{sub:an_dom}.
Lastly, in Section \ref{sub:an_conclusion} we combine all the results to conclude that the algorithm outline proves Conjecture \ref{conjecture:3_5}.

\subsection{A policy for ensuring high average gain}
\label{sub:an_gain}

We have seen that it suffices to guarantee an average gain of $5$ points per move (Claim \ref{claim:avg_5}), 
and also that the last move on any component gains at least $5$ points (Corollary \ref{cor:last_5}).
Therefore, it suffices to guarantee that each pair of consecutive Dominator and Staller moves gains at least $10$ points in order to make sure that Dominator wins the game.

\begin{definition}
\label{def:excess_gain}
Define the \emph{excess gain} of move $m_t$ at step $t$, denoted by $\psi_t$, as follows.
\begin{itemize}
	\dnsitem If $t$ is odd (Dominator plays $m_t$), then $\psi_t = g(m_t, t) - 7$.
	\dnsitem If $t$ is even (Staller plays $m_t$), then $\psi_t = g(m_t, t) - 3$.
\end{itemize}
Additionally, define the \emph{cumulative excess gain} at step $t$ to be the sum of excess gains in steps $1$ through $t$, and denote it by 
$$\Psi_t = \sum_{i=1}^{t} \psi(m_{i}, i).$$
\end{definition}

\begin{observation}
\label{obs:sufficient_overhead}
Each of the following conditions is sufficient in order for Dominator to win.
\begin{enumerate}
	\dnsitem $T \leq T_{\max}$.
	\dnsitem $\Psi_T \geq 0$.
	\dnsitem $T$ is odd (i.e., Dominator plays $m_T$), and $\Psi_T \geq -2$.
	\dnsitem $T$ is even (i.e., Staller plays $m_T$), and $\Psi_{T - 1} \geq -2$. 
\end{enumerate}
\end{observation}
\Proof
$ $ 
\begin{enumerate}
	\dnsitem $T \leq T_{\max}$ is the condition in Conjecture \ref{conjecture:3_5}.

	\dnsitem If $\Psi_T \geq 0$, then the game ended with an average gain of at least $5$ points, which is a sufficient condition according to Claim \ref{claim:avg_5}.
	
	\dnsitem If $T$ is odd and $\Psi_T \geq -2$, then $T - 1$ is even and $5 \cdot (T - 1) + 7 - 2 = 5 \cdot T$ points are gained in $T$ moves. Therefore an average gain of $5$ points is achieved throughout the game, which is a sufficient condition by Claim \ref{claim:avg_5}.
	
	\dnsitem If $T$ is even and $\Psi_{T - 1} \geq -2$ then $\Psi_T \geq 0$, since Corollary \ref{cor:last_5} guarantees that $\psi_T \geq 2$. Therefore this is a sufficient condition.
\QED
\end{enumerate}

\paragraph{}
The following guarantees are maintained throughout the execution, as will be shown in the analysis.
\begin{enumerate}
	\dnsitem Every Staller move gains at least $3$ points, i.e., $\psi_{2t} \geq 0$ for all $t$.
	\dnsitem $\Psi_t \geq -2$ for all $1 \leq t \leq T$.
	\dnsitem The algorithm never relies on past gains, but rather on future gains. Namely, it guarantees that if $\psi_t(v)$ is negative at some point, then there will be positive excess gain in future moves to make up for it.
Therefore, if $\Psi_t > 0$ for some $t$, we can use the cumulative excess gain to convert $B_2$ vertices to $B_3$.

\end{enumerate}

\subsection{Preliminary properties simplifying the analysis}
\label{sub:an_simplify}

In this section we prove properties which will allow us to calculate a lower bound for the gain of playing a (real) vertex from the dense graph directly on the box containing it in the dense graph.
First, we describe the difference between playing on the dense graph and playing on the underlying graph.
Then, we prove that it suffices to analyze moves on the dense graph by analyzing them directly on the box that contains them.
Since Dominator moves are chosen from vertices of the dense graph, this includes all Dominator moves and all Staller moves that are on the dense graph, i.e., all moves that are not under triplet subtrees.
Staller moves that are not on the dense graph will be analyzed separately in Section \ref{sub:an_staller}.

\begin{figure}[thbp]
  \caption{\sf The graphs of Lemma \ref{claim:ana_dense}.}
  \medskip
  \centering
    \fbox{\includegraphics[width=7cm]{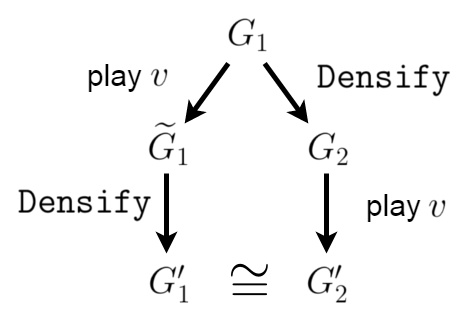}}
  \label{fig:dense_graph_claim}
\end{figure}

\begin{lemma}
\label{claim:ana_dense}
Let $G_1$ be a graph, and let $G_2$ be the corresponding dense graph.
Let $v$ be a (real) vertex which exists both in $G_1$ and in $G_2$.
If $g$ points are gained when $v$ is played directly on $G_2$ (i.e., without invoking $\densify()$) and the resulting graph (again, without invoking $\densify()$) is $G'_2$, 
then $g$ points can be gained when $v$ is played on $G_1$, yielding $\widetilde{G}_1$, and the dense graph $G'_1$ resulting from $\densify(\widetilde{G}_1)$ is the same as $G'_2$ except that it may have (at most three) additional $B_2W$ components.
See Figure \ref{fig:dense_graph_claim} for an illustration of the graphs in the claim.
\end{lemma}
\Proof
Consider $G_1$, $G_2$ and $v$ as in the claim. 
Since $G_2 = \densify(G_1)$, the only difference between the graphs $G_1$ and $G_2$ is the replacement of triplet subtrees from $G_1$ with white leaves in $G_2$.
Therefore, if $G_1$ does not contain triplet vertices, then $G_1 = G_2$.
Assume that $G_1$ contains at least one triplet vertex, and let $u$ be a triplet head (guaranteed to exist by Claim \ref{claim:triplet_vertex_implies_head}).
Let $u_1, u_2$ and $u_3$ be the triplet witnesses of $u$ in $G_1$, and let $\lambda$ be the white leaf adjacent to $u$ in $G_2$ that is not in $G_1$ 
(the procedure $\densify$ guarantees that exactly one such leaf exists, since it creates a single virtual leaf next to each triplet head).
Denote by $T_u$ the set of vertices in the triplet subtree rooted at $u$ in $G_1$, excluding $u$.
Observation \ref{obs:triplet_subtrees_white} guarantees that all vertices in $T_u$ are white, and we know that $\lambda$ is white.
$\lambda$ is not in $G_1$, and all vertices of $T_u$ are not in $G_2$, therefore by the claim's assumption, $v$ cannot be any of these vertices.

If $v \neq u$, then the above implies that all vertices of $T_u$ (in $G_1$) and $\lambda$ (in $G_2$) remain white. Therefore $u$ is a triplet head in $\widetilde{G}_1$ with the same triplet witnesses (this results from the way we choose triplet witnesses when there are more than three potential witnesses), which implies that $u$ has one virtual (white) leaf neighbor in $G'_1 = \densify(\widetilde{G}_1)$.
Note that $u$ itself has the same color (and value) in $G'_1$ and in $G'_2$.

\begin{figure}[thbp]
  \caption{\sf (a) An example triplet subtree rooted at a vertex $u$ in $G_1$, with triplet witnesses $u_1, u_2$ and $u_3$. All unlabeled vertices are white.
		(b) The components of $\widetilde{G}_1$ resulting from the triplet subtree following the move $v = u$.}
  \medskip
  \centering
    \fbox{\includegraphics[width=9cm]{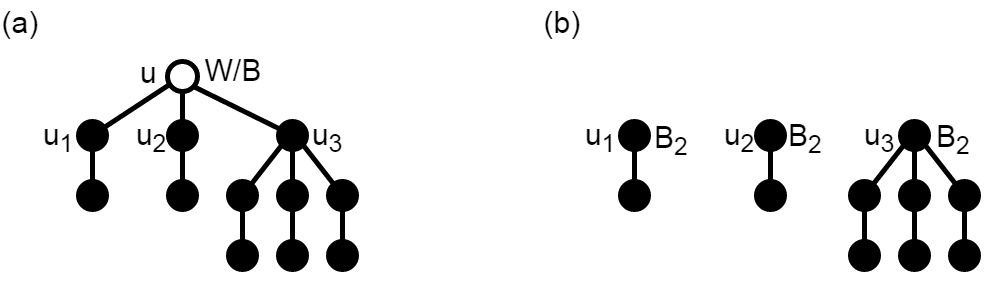}}
  \label{fig:triplet_subtree_move}
\end{figure}

If $v = u$, then $\lambda$ becomes red when $v$ is played on $G_2$, and adds exactly $3$ points to the gain (since it is white in $G_2$).
In $G_1$, each of the triplet witnesses $u_1, u_2$ and $u_3$ becomes a blue vertex in a separate component, and may be set as a $B_2$ or $B_3$ vertex.
For each $1 \leq i \leq 3$, if $u_i \in \mathcal{WT}_2$ in $G_1$, then $u_i$ is in a $BW$ component in $\widetilde{G}_1$, and otherwise (i.e., if $u_i$ is a triplet vertex in $G_1$), $u_i$ is a blue triplet head of degree $3$ in a component of $\widetilde{G}_1$. See Figure \ref{fig:triplet_subtree_move} for an example.
In both cases, $u_i$ is a blue vertex in a $BW$ component in $G'_1$, and can be converted to $B_2$ without violating Invariant $\invboxes$.
Therefore exactly $3$ points can be gained from the vertices of $T_u$.

The above shows that each virtual leaf in $G'_1$ has a corresponding white leaf in $G'_2$, except possibly for three $BW$ components, and that each white leaf in $G'_2$ has a corresponding white leaf in $G'_1$.
The claim follows from the fact (mentioned above) that all other vertices (except for the vertices in the three discussed $BW$ components) are the same in both graphs.
\QED

We now prove that in some cases, it is possible to calculate a lower bound for the gain of a move $m_t$ by calculating its gain on another, simpler graph.
Specifically, this simpler graph can be any graph that contains only the vertices of the box $Q$ that contains $m_t$, in some valid box decomposition $\mathcal{Q}$ of $\hat{G}_{t-1}$.
We use this important property later in the analysis of Dominator and Staller moves.

\begin{lemma}
\label{claim:ana_box}
Suppose $\hat{G}_{t-1}$ is good.
Let $v$ be a vertex in a box $Q$ of a valid box decomposition $\mathcal{Q}$ of $\hat{G}_{t-1}$ such that all boxes in $\mathcal{Q} \setminus \left\{Q\right\}$ are not corrupted, 
and consider the graph $G'_{t-1}$ that contains a single component with a box decomposition $\mathcal{Q}'$ containing a single box $Q'$ that is identical to $Q$.
Suppose $g$ points can be gained by playing $v$ in $G'_{t-1}$ and the resulting underlying graph $G'_t$ is good. 
Then it is possible to play $v$ in $\hat{G}_{t-1}$ so that it leads to a good $\hat{G}_t$.
More precisely, the following properties hold.
\begin{description}
	\dnsitem[$\boxpropnocorhigh$.]
	If $G'_t$ does not contain a semi-corrupted component, and one or more of the following conditions is satisfied:
	\begin{enumerate}
		\dnsitem 
		\label{claim:ana_box:nocorhigh:rootb}
		$Q$ is a root box in $\hat{G}_{t-1}$.
		\dnssubitem 
		\label{claim:ana_box:nocorhigh:boxr}
		$v$ is the box root of $Q$.
		\dnssubitem 
		\label{claim:ana_box:nocorhigh:nodense}
		The box root $r$ of $Q$ does not exist in $\hat{G}_t$ (i.e., $r$ becomes red. Note that this may occur even if $r \neq v$).
		\dnssubitem 
		\label{claim:ana_box:nocorhigh:yesroot}
		The box root $r$ of $Q'$ exists in $G'_t$ (i.e., $r$ does not become red).
	\end{enumerate}
	Then at least $g$ points are gained by playing $v$ in $\hat{G}_{t-1}$,
	 and the resulting graph $\hat{G}_t$ is good (namely, it has a valid box decomposition) and does not contain semi-corrupted components.
	
	\dnsitem[$\boxpropnocorlow$.]
	If $G'_t$ does not contain a semi-corrupted component, and none of the above conditions (\ref{claim:ana_box:nocorhigh:rootb}) - (\ref{claim:ana_box:nocorhigh:yesroot}) are satisfied,
	then at least $g - 3$ points are gained by playing $v$ in $\hat{G}_{t-1}$, 
	and the resulting graph is good and does not contain semi-corrupted components.

	\dnsitem[$\boxpropcor$.]
	If $G'_t$ contains a semi-corrupted component $C'$ and one or more of the following conditions holds:
	\begin{enumerate}
		\dnsitem The corrupted box of $C'$ (under some valid box decomposition) is a root box of (a valid box decomposition of) $\hat{G}_t$. Note that this includes the case that $Q$ is a root box in $\hat{G}_{t-1}$.
		\dnssubitem Playing $r$ on $C'$, where $r$ is the box root of $Q$, gains at least $8$ points in $G'_t$, and the resulting graph is good.
			\end{enumerate}
	Then at least $g$ points are gained by playing $v$ in $\hat{G}_{t-1}$,
	 and the resulting graph is good (and may contain up to one semi-corrupted component).
\end{description}
\end{lemma}
Note that the classification of Lemma \ref{claim:ana_box} does not cover all cases in which $G'_t$ contains a semi-corrupted component, because not all cases are needed for the rest of the analysis.
\Proof
Assume all conditions of the claim are satisfied, and let $\mathcal{Q}_1$ be a valid box decomposition of $G'_t$.
Consider an intermediate partition $\mathcal{P}$ of the vertices of $\hat{G}_{t}$ which is constructed by the following procedure.

\smallskip
\par\noindent{\bf Procedure $\interpart$}:
\begin{enumerate}
	\dnsitem For each box $Q_i \in \mathcal{Q}$ whose vertices exist in $\hat{G}_{t}$ (i.e., such that none of its vertices become red in move $m_{t}$),
	add $Q_i$ to $\mathcal{P}$.
	Note that this includes all boxes of $\mathcal{Q}$ excluding $Q$, and possibly excluding its parent box $P$ (if exists) 
	as well as some high leftover boxes of size $1$ from $\mathcal{Q}$ whose root (and only) vertex becomes red.
	\dnsitem Add all boxes of $\mathcal{Q}_1$ to $\mathcal{P}$.

	\dnsitem If $Q$ is not a root box in $\hat{G}_{t-1}$, and its box root $r$ is in $\hat{G}_{t}$ but not in $G'_{t}$,
	then add the box $Q_r$ of size $1$ containing $r$ to $\mathcal{P}$.
	Note that $Q_r$ is either high leftover (if $r$ is high) or corrupted (otherwise).

	\dnsitem If $Q$ is not a root box in $\hat{G}_{t-1}$, and not all vertices of its parent box $P$ are in $\hat{G}_t$, 
	then add the sets $Q_i$ containing all maximal connected subsets of $P$ to $\mathcal{P}$.
	\dnsitem For every component $C_1$ of size $2$ in $\hat{G}_{t}$, 
	add the box $P_1$ containing all vertices of $C_1$ to $\mathcal{P}$ (if it is not already in $\mathcal{P}$). 
\end{enumerate}
Note that $\mathcal{P}$ is not necessarily a valid box decomposition.
Also observe that $\mathcal{P}$ is a partition of the vertices of $\hat{G}_{t}$, and that all the sets added to $\mathcal{P}$ by the above five steps are boxes.
After performing Procedure $\interpart$, we invoke Operation $\discblue$ on the resulting graph $\hat{G}_{t}$ and disconnect all external edges in $\mathcal{P}$ that connect two blue vertices.
\smallskip
\par\noindent
{\bf Operation $\discblue$:}
\begin{itemize}
\dnsitem[] For every edge $e = (u_1, u_2)$ that is external in $\mathcal{P}$, do:
	\begin{itemize}
		\dnsitem[] If $u_1$ and $u_2$ are both blue, remove $e$.
	\end{itemize}
\end{itemize}
\bigskip
Note that after performing Operation $\discblue$ there are no parent boxes of size $1$ in $\mathcal{P}$ (since box roots are blue).

We now consider the partition $\mathcal{P}$ under the three different settings specified in the claim, and describe in each setting how $\mathcal{P}$ can be modified into a valid box decomposition of $\hat{G}_t$, so $\hat{G}_t$ is good, while achieving the desired properties (i.e., a gain of at least $g$ points with no corrupted boxes in Case $\boxpropnocorhigh$, a gain of at least $g - 3$ points with no corrupted boxes in Case $\boxpropnocorlow$, and a gain of at least $g$ points with at most one semi-corrupted component in Case $\boxpropcor$). This will imply the claim.
We rely on the following three claims.

\begin{claim}
\label{claim:ana_box:p1_bw}
All parent boxes of size $2$ in $\mathcal{P}$ are of the form $BW$.
\end{claim}
\Proof
Let $P_1$ be a parent box of size $2$ in $\mathcal{P}$, 
and assume towards contradiction that $P_1$ is of the form $WW$.
Recall that Observation \ref{obs:w_neighborhood} guarantees that all neighbors of each white vertex are still in the graph.
Also recall that Definitions \ref{def:box_general} and \ref{def:box_decomp} guarantee that each box that is not a root box contains a blue vertex (the box root), and that a parent box is of size $3$ or more (Property $\decompsize$).
Since $P_1$ is in $\mathcal{P}$ and $\mathcal{P}$ was constructed using Procedure $\interpart$, we conclude that $P_1$ was added to $\mathcal{P}$ for one of the following three reasons.
The first option is that $P_1$ was in $\mathcal{Q}$. This is impossible, since then $P_1$ would be a parent box of size $2$, which contradicts Property $\decompsize$ of Definition \ref{def:box_decomp}.
The second option is that $P_1$ is in $\mathcal{Q}_1$. This is impossible when $P_1$ is white, since a box of the form $WW$ cannot be a parent box and cannot have a parent, which means that it must be in a component of size $2$ in $G'_t$. This contradicts the fact that the component containing $P_1$ in $G'_t$ must contain a blue vertex by Observation \ref{obs:w_neighborhood}.
The third option is that $P_1 \subseteq P$, where $P$ is the parent box of $Q$ in $\hat{G}_{t-1}$. This is also impossible when $P_1$ is white since, as in the previous case, $P$ must contain a blue vertex by Observation \ref{obs:w_neighborhood}.
The claim follows.
\QED

\begin{claim}
\label{claim:ana_box:p1_root}
All boxes of size $2$ in $\mathcal{P}$ are root boxes.
\end{claim}
\Proof
Let $P_1$ be a box of size $2$ in $\mathcal{P}$.
Denote by $u_1$ and $u_2$ the blue and white vertices of $P_1$, respectively.
If $P_1$ was a root box in $\hat{G}_{t-1}$, then it is clearly also a root box in $\hat{G}_t$.
Otherwise, we know that $u_1$ was not a box root:
Assume towards contradiction that $u_1$ was a box root in $\hat{G}_{t-1}$, and denote the box which contained $u_1$ and $u_2$ in $\hat{G}_{t-1}$ by $Q_0$.
Definition \ref{def:box_general} guarantees that box roots do not have neighbors that are white leave, therefore $u_2$ was not a leaf in $Q_0$, and had a neighbor, $u_3$.
Since $u_2$ is white in $\hat{G}_t$, we conclude from Observation \ref{obs:w_neighborhood} that $u_3$ is in $\hat{G}_t$ (and Procedure $\interpart$ guarantees that it is in $P_1$), in contradiction to the assumption that $P_1$ is of size $2$.
Therefore $u_1$ was not a box root in $\mathcal{Q}$, which means that it does not have a parent in $\mathcal{P}$.
In all cases, we conclude that $P_1$ is a root box in $\mathcal{P}$.
\QED

\begin{claim}
\label{claim:ana_box:fix_p}
Assume that $g_0$ points are gained, and the described partition $\mathcal{P}$ 
does not contain corrupted boxes and satisfies all properties of Definition \ref{def:box_decomp} except, possibly, for Properties $\decompsize$ and $\decompcleanl$.
Then $G_t$ is good, and $\mathcal{P}$ can be modified into a valid box decomposition of $\hat{G}_t$ that does not contain corrupted boxes, while gaining at least $g_0$ points.
\end{claim}
\Proof
We start by handling the case that $\mathcal{P}$ does not satisfy Property $\decompcleanl$, i.e., it contains a high box that is the parent of at least one high leftover box.
Consider the following operation.
\smallskip
\par
\begin{figure}[thbp]
  \caption{\sf Examples of the results of Operation $\fixbwparent$ for each pair of boxes $Q_1$ and $P_1$.
		(a) The resulting box $Q_2$, if $Q_1$ is dispensible of type $1$ or high leftover.
		(b) An example of the resulting boxes $Q_3$ and $Q_4$ when $Q_1$ is dispensible of type $2$.}
  \medskip
  \centering
    \fbox{\includegraphics[width=9cm]{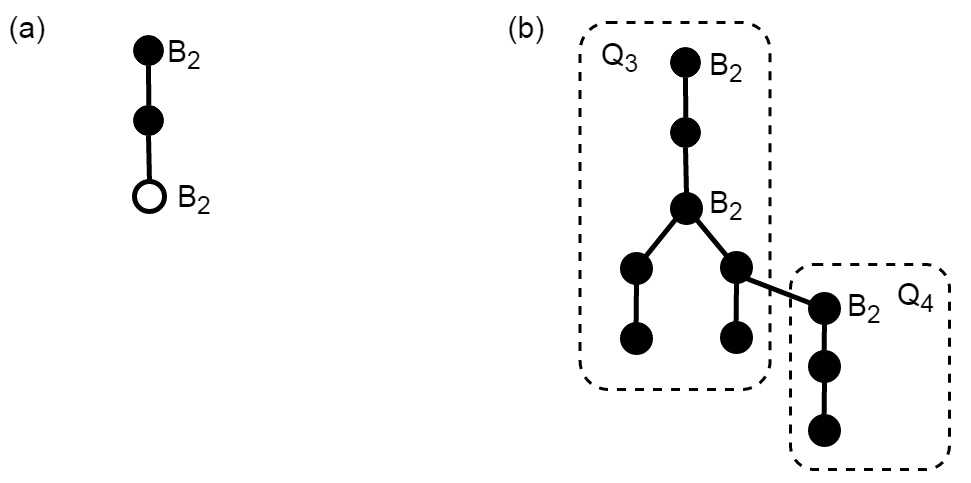}}
  \label{fig:ana_box_to_graph}
\end{figure}

\noindent
{\bf Operation $\joinhigh$:}
\begin{itemize}
\dnsitem[] While there are high boxes $P_1$ and $Q_1$ such that $P_1$ is the parent of $Q_1$, do:
	\begin{itemize}
		\dnsitem[] Remove both boxes $P_1$ and $Q_1$ from $\mathcal{P}$ and replace them with the single box $Q_2$ that contains all vertices of $Q_1$ and $P_1$.
	\end{itemize}
\end{itemize}
\bigskip
After performing Operation $\joinhigh$, $\mathcal{P}$ satisfies Property $\decompcleanl$, and the gain does not change.
It is possible that now $\mathcal{P}$ is a valid box decomposition of $\hat{G}_t$.
If this is not the case, it remains to modify $\mathcal{P}$ so that it satisfies Property $\decompsize$.
From Claims \ref{claim:ana_box:p1_bw} and \ref{claim:ana_box:p1_root} we conclude that all parent boxes of size $2$ in $\mathcal{P}$ are root boxes of the form $BW$, and we have seen earlier that after performing Operation $\discblue$ $\mathcal{P}$ does not contain parent boxes of size $1$.
Let $P_1$ be a (root) parent box of size $2$ in $\mathcal{P}$, and let $Q_1$ be another box in $\mathcal{P}$ such that $P_1$ is the parent of $Q_1$.
Consider the following operation. See Figure \ref{fig:ana_box_to_graph} for illustrations.

\smallskip
\par\noindent
{\bf Operation $\fixbwparent$:}
\begin{itemize}
\dnsitem[] For every parent box $P_1 = (u_1, u_2)$ in $\mathcal{P}$, such that $u_1$ is blue and $u_2$ is white, do:
	\begin{enumerate}
		\dnsitem Find a box $Q_1$ in $\mathcal{P}$ such that $P_1$ is the parent of $Q_1$, and denote the box root of $Q_1$ by $r_1$.
		\dnsitem Remove both boxes $P_1$ and $Q_1$ from $\mathcal{P}$ and replace them with the single box $Q_2$ that contains all vertices of $Q_1$ and $P_1$.
		\dnsitem Convert $u_1$ and $r_1$ to $B_2$ (if they are high).
		\dnsitem If $Q_1$ was a dispensible box of type $2$, then:
		\begin{enumerate}
			\dnsitem Denote by $r_2$ the $B_2$ vertex of $Q_1$ that is not $r_1$, and denote by $u_3$ and $u_4$ the two vertices in $Q_1$ that are on its high subtail.
			\dnssubitem Remove $Q_2$ from $\mathcal{P}$, and replace it with the following two boxes:
			The box $Q_3$, which contains all vertices of $Q_2$ except for $r_2, u_3$ and $u_4$, 
			and the box $Q_4$, which contains $r_2$, $u_3$ and $u_4$ with box root $r_2$. 
		\end{enumerate}
	\end{enumerate}
\end{itemize}
\bigskip

Observe that Operation $\fixbwparent$ may only increase the gain (if $u_1$ or $r_1$ is high).
We make the following two observations.
First, if $Q_1$ was a dispensible box of type $1$ or a high leftover box, then $Q_2$ is a regular colored box, since it satisfies Property $\propbbr$:(\ref{invnp:2tail}) of Definition \ref{def:regular_colored}
(see Case (a) in Figure \ref{fig:ana_box_to_graph}).
Second, if $Q_1$ was a dispensible box of type $2$, then the box $Q_3$ is a regular colored box (again by Property $\propbbr$:(\ref{invnp:2tail})), and the box $Q_4$ is dispensible of type $1$ (see an example in Case (b) of Figure \ref{fig:ana_box_to_graph}).
We conclude that after performing Operation $\fixbwparent$, the resulting $\mathcal{P}$ satisfies Definition \ref{def:box_decomp} and does not contain corrupted boxes, and therefore it is a valid box decomposition of $\hat{G}_t$, and that at least $g_0$ points are gained.
\QED

For Cases $\boxpropnocorhigh$ and $\boxpropnocorlow$, it remains to show that in each setting, the gain is as described in the theorem and $\mathcal{P}$ satisfies Properties $\decomponereg$, $\decomponecorr$, $\decomproot$ and $\decompedges$ of Definition \ref{def:box_decomp} before performing Operations $\joinhigh$ and $\fixbwparent$, and then the theorem will follow from Claims \ref{claim:ana_box:p1_root} and \ref{claim:ana_box:fix_p}.
In Case $\boxpropcor$, we cannot use Claim \ref{claim:ana_box:fix_p} directly.

\bigskip
\par\noindent
{\bf Case ($\boxpropnocorhigh$):}
Let us first consider the setting of Case ($\boxpropnocorhigh$), in which $G'_t$ does not contain a semi-corrupted component, and one of the four conditions (1)-(4) is satisfied.
We examine each of these conditions.
\smallskip
\par\noindent
{\bf Subcase ($\ref{claim:ana_box:nocorhigh:rootb}$):}
	First, suppose Condition (\ref{claim:ana_box:nocorhigh:rootb}) holds, namely, $Q$ is a root box in $\hat{G}_{t-1}$. 
	Since box roots are always blue, external edges always connect a box root to its parent, and root boxes do not have parents (see Definitions \ref{def:box_general} and \ref{def:box_decomp}),
	we conclude that all external neighbors of vertices of $Q$ (i.e., neighbors that are in $\hat{G}_{t-1}$ but not in $Q$) are blue box roots.
	This implies that if a vertex in $Q$ does not have internal white neighbors (i.e., white neighbors that are also in $Q$), then it does not have any white neighbor.
	We conclude that $Q$ does not contain vertices that are in $\hat{G}_t$ but are not in $G'_t$ (i.e., vertices that are red in $G'_t$ but blue in $\hat{G}_t$). 
	Additionally, if an external neighbor $u$ does not have internal white neighbors in its box in $\mathcal{Q}$, then it must be in a high leftover box of size $1$, 
	which means (since box roots cannot be parents) that it does not have additional external neighbors except its parent.
	Therefore, the only case in which an external neighbor's color changes as a result of playing $v$ is when a vertex in a high leftover box of size $1$ becomes red, and in such a case $3$ more points are gained and no additional boxes are modified.
	
	We make the following four observations regarding the properties of Definition \ref{def:box_decomp}.
	First, each connected component contains at most one regular box, i.e., Property $\decomponereg$ is satisfied.
	This is because all boxes from $\mathcal{Q}$ that were not in the same component with $v$ are in $\mathcal{P}$, 
	all boxes of $\mathcal{Q} \setminus \{Q\}$ that were in the same component with $v$ were not regular,
	and $\mathcal{Q}_1$ is a valid box decomposition and therefore satisfies this condition.	
	Second, $\mathcal{P}$ does not contain corrupted boxes, and therefore Property $\decomponecorr$ is satisfied.
	This is because the only potentially corrupted box is $Q_r$, generated in step $3$ of Procedure $\interpart$, and from the previous paragraph we conclude that it does not exist when $Q$ is a root box. 
	Third, Property $\decomproot$ holds, since all box roots have at most one external neighbor, and if such a neighbor exists then it is not another box root.
	This is because all external edges connecting blue vertices were disconnected in Operation $\discblue$ (and box roots are blue).
	Fourth, all external edges connect a box root to its parent, 	
	because all external edges of $\mathcal{P}$ are either external edges in $\mathcal{Q}_1$, or were external edges in $\mathcal{Q}$.  
	Therefore Property $\decompedges$ holds.

	We conclude from Claim \ref{claim:ana_box:fix_p} that $\mathcal{P}$ can be modified into a valid box decomposition of $\hat{G}_t$ while preserving a gain of $g$ points or more, and that $\hat{G}_t$ does not contain semi-corrupted components, which is what we wanted to prove.
	
\smallskip
\par\noindent
{\bf Subcase ($\ref{claim:ana_box:nocorhigh:boxr}$):}
	Next, suppose Condition ($\ref{claim:ana_box:nocorhigh:boxr}$) of Case ($\boxpropnocorhigh$) holds, namely,
	$Q$ is not a root box, and $v$ is the box root of $Q$.
	Since only the box root $v$ had a parent in $\mathcal{Q}$, and $v$ is red, we conclude that all boxes from $\mathcal{Q}_1$ are root boxes in $\mathcal{P}$.
	Therefore all components except, possibly, for components which contain vertices from the parent box $P$, 
	satisfy Properties $\decomponereg$, $\decomponecorr$, $\decomproot$ and $\decompedges$, for the same reasons as in the previous subcase.
	It remains to handle the components that contain vertices from $P$.
	Consider the boxes $P_i$ in $\mathcal{P}$ which resulted from the parent box $P$ in $\mathcal{Q}$.
	Exactly one of the following cases occurs.
	\begin{enumerate}
		\dnsitem All vertices of $P$ are in $\hat{G}_t$.
		Then the box $P$ remains as it was, except that maybe the vertex $p$ that was the parent of the box root $r$ of $Q$ was converted from white to blue.
		Whether $P$ was a regular, dispensible or high leftover box, it is still of the same type that it was, since all these properties are still satisfied if a single vertex is converted from white to $B_3$ (see Section \ref{sub:box_types}). 
		Therefore in this case $\mathcal{P}$ satisfies Properties $\decomponereg$, $\decomponecorr$, $\decomproot$ and $\decompedges$. 
		
		\dnsitem Some vertex of $P$ became red and is not in $\hat{G}_t$.
		Then at least $2$ additional points were gained, and they can be used to convert all remaining vertices of $P$ to high vertices (since each box contains at most two $B_2$ vertices).
		Claim \ref{claim:ana_box:p1_root} guarantees that all resulting boxes of size $2$ are root boxes, and we conclude that all properties of Definition \ref{def:box_decomp} are satisfied on these components as well.
	\end{enumerate}

	From Claim \ref{claim:ana_box:fix_p}, we conclude that $\mathcal{P}$ can be modified into a valid box decomposition of $\hat{G}_t$ while preserving a gain of $g$ points or more, and that $\hat{G}_t$ does not contain semi-corrupted components.

\smallskip
\par\noindent
{\bf Subcase ($\ref{claim:ana_box:nocorhigh:nodense}$):}
	Next, consider Condition ($\ref{claim:ana_box:nocorhigh:nodense}$) of Case ($\boxpropnocorhigh$), namely,
	the box root $r$ of $Q$ does not exist in $\hat{G}_t$ (i.e., $r$ becomes red), and assume that $v \neq r$.
	Then all boxes of $\mathcal{Q}_1$ are root boxes in $\mathcal{P}$ (since only the box root $r$ could have a parent), and all components that do not contain vertices from $\mathcal{Q}_1$ remain as they were in $\hat{G}_{t-1}$.
	As in the previous subcase, 
	we conclude that Properties $\decomponereg$, $\decomponecorr$, $\decomproot$ and $\decompedges$ are satisfied by $\mathcal{P}$, and therefore from Claim
	\ref{claim:ana_box:fix_p} we conclude that $\hat{G}_t$ is good and does not contain semi-corrupted components, and at least $g$ points are gained.
	
\smallskip
\par\noindent
{\bf Subcase ($\ref{claim:ana_box:nocorhigh:yesroot}$):}
	Finally, suppose Condition ($\ref{claim:ana_box:nocorhigh:yesroot}$) of Case ($\boxpropnocorhigh$) holds, i.e.,
	$Q$ is not a root box and the box root $r$ of $Q'$ exists in $G'_t$ (that is, it does not become red). Denote by $Q'_1$ the box containing $r$ in $\mathcal{Q}_1$.
	Since box roots do not have internal neighbors that are white leaves, and Observation \ref{obs:w_neighborhood} guarantees that the neighborhood of a white vertex remains as it was, 
	we conclude that $Q'_1$ is not of size $2$.
	Therefore Properties $\decomponereg$ and $\decompedges$ are satisfied in $\mathcal{P}$. 
	Properties $\decomponecorr$ and $\decomproot$ are satisfied for the same reasons as in Subcase ($\ref{claim:ana_box:nocorhigh:rootb}$).
	We conclude from Claim \ref{claim:ana_box:fix_p} that $\hat{G}_t$ has a valid box decomposition that does not contain corrupted boxes, and that at least $g$ points are gained.

\smallskip
\par\noindent
In all the above subcases, the gain in $\hat{G}_t$ is at least as high as the gain in $G'_t$ and the resulting graph is good (with no semi-corrupted components).
Hence Case ($\boxpropnocorhigh$) follows.

\bigskip
\par\noindent
{\bf Case ($\boxpropnocorlow$):}
We now turn to Case $(\boxpropnocorlow)$ of the claim, in which $Q$ is not a root box, and the box root $r$ is in $\hat{G}_t$ but not in $G'_t$.
Consider the following operation.
\smallskip
\par\noindent
{\bf Operation $\convhigh$:}
\begin{itemize}
\dnsitem[] Convert the root box $r$ of $Q$ to $B_3$.
\end{itemize}
The difference between the gain in $G'_t$, and the gain in $\hat{G}_t$ after performing Operation $\convhigh$ (if $r$ was $B_2$), is at most $3$ points.
Since $r$ was a box root and box roots cannot be parents, the box $Q_r$ containing $r$ in $\mathcal{P}$ is a high leftover box of size $1$ that is not the parent of another box, 
and its parent box $P$ remains as it was in $\hat{G}_{t-1}$.
We conclude that the component $C$ containing $r$ satisfies all properties of Definition \ref{def:box_decomp} except, possibly, for Property $\decompcleanl$ (i.e., $P$ may also be a high box).
All other components can be analyzed as in Case ($\boxpropnocorhigh$) above, and therefore we conclude from Claim \ref{claim:ana_box:fix_p} that $\mathcal{P}$ can be converted into a valid box decomposition of $\hat{G}_t$ while gaining at least $g - 3$ points, and that $\hat{G}_t$ does not contain semi-corrupted components.

\bigskip
\par\noindent
{\bf Case ($\boxpropcor$):}
Finally, we consider Case $(\boxpropcor)$. 
If $G'_t$ contains a semi-corrupted component $C'$ and one of the specified conditions holds, then at least one of the conditions of Case $\boxpropnocorhigh$ above is satisfied (except that now one of the resulting components may contain a corrupted box).
All components except, possibly, for the component $C^*$ containing vertices from $C'$, satisfy the conditions of Claim \ref{claim:ana_box:fix_p}, and therefore $\mathcal{P}$ can be modified so that its restriction to these components is valid, while preserving the number of points gained on them.
It remains to handle $C^*$.
Property $\decompsize$ of Definition \ref{def:box_decomp} guarantees that the root box of $\mathcal{Q}_1$ that is in $C'$ is a box $Q_{C'}$ of size at least $3$ in $\mathcal{P}$,
therefore Property $\decompsize$ is also satisfied by $\mathcal{P}$. 
Since $\mathcal{P}$ contains a single corrupted box, we conclude that $\mathcal{P}$ satisfies Properties $\decomponereg$, $\decomponecorr$, $\decomproot$ and $\decompedges$ on $C^*$ for the same reasons as in Case ($\boxpropnocorhigh$) above.
Therefore, after performing Operation $\joinhigh$, $\mathcal{P}$ is a box decomposition of $\hat{G}_t$ and at least $g$ points are gained. 

It remains to check whether $C^*$ is semi-corrupted.
In order for $C^*$ to be semi-corrupted (and for $\mathcal{P}$ to be a valid box decomposition of $\hat{G}_t$), we need to check if it contains a move $u$ gaining at least $8$ points with a good resulting graph $\hat{G}_{t+1}$.
Since $C'$ is semi-corrupted, it contains a move $u$ gaining at least $8$ points when played in $G'_t$.
Under each of the conditions of the claim, one of the conditions of Case $\boxpropnocorhigh$ is satisfied for step $t+1$ (since the move $u$ in question is either a box root, or on a root box), and we conclude that at least $8$ points can be gained when playing on $\hat{G}_{t}$ so that $\hat{G}_{t+1}$ is good. Therefore $C^*$ is semi-corrupted, as desired.

This concludes the proof of Lemma \ref{claim:ana_box}.
\QED

\subsection{Two special subtrees}
\label{sub:special_subtrees}

In this section we focus attention on two types of subtrees that occur frequently in subsequent analysis, 
and describe how the algorithm may cope with such subtrees, and what moves can be used in the analysis.
We start by defining the subtrees.

\begin{figure}[thbp]
  \caption{\sf An example of a fix vertex (marked as $u$) and its fix subtree.}
  \medskip
  \centering
    \fbox{\includegraphics[height=1.5cm]{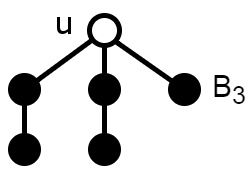}}
  \label{fig:tu_vertex}
\end{figure}
\begin{definition}
A vertex $u \in V$ is called a \emph{fix} vertex if the following conditions hold.
\begin{enumerate}
	\dnsitem $u$ has (at least) one neighbor that is a $B_3$ leaf.
	\dnsitem $u$ has (at least) two high tails of length $2$.
	\dnsitem There is a valid box decomposition where none of the vertices in these tails are box roots. 
		Note that this implies that $u$ is white.
\end{enumerate}
If $u$ has at most one additional white neighbor, and this neighbor (if exists) is not the lead of a white tail of length $1$ or $2$, 
we say that $u$ is a \emph{strong fix} vertex. 

The subtree containing $u$ and the three specified tails is called the \emph{fix subtree rooted at $u$}.
See Figure \ref{fig:tu_vertex}.
\end{definition}

\begin{figure}[thbp]
  \caption{\sf Semi-triplet split vertices (marked as $u$) and their semi-triplet subtrees.}
  \medskip
  \centering
    \fbox{\includegraphics[width=8cm]{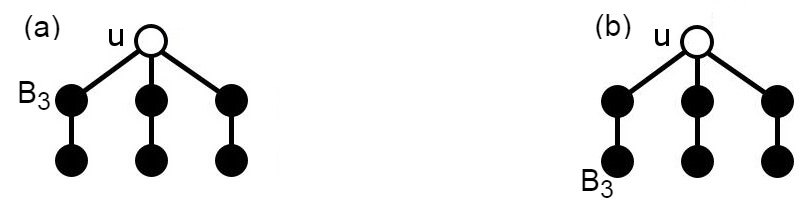}}
  \label{fig:semi_triplet_split}
\end{figure}
\begin{definition}
A high split vertex $u$ is called a \emph{semi-triplet} vertex if the following conditions hold. 
\begin{enumerate}
	\dnsitem $u$ has three high tails of length $2$, and there is a valid box decomposition where none of the vertices in these tails are box roots. 
	\dnsitem $u$ is not a triplet vertex (i.e., not all vertices in the tails are white).
\end{enumerate}

If $u$ has at most one additional white neighbor, and this neighbor (if exists) is not the lead of a white tail of length $1$ or $2$, 
we say that $u$ is a \emph{strong semi-triplet} vertex.

The subtree containing $u$ and the three specified tails is called the \emph{semi-triplet subtree rooted at $u$}. 
See Figure \ref{fig:semi_triplet_split}.
\end{definition}

\begin{lemma}
\label{lemma:fix_subtree_gain}
Let $Q$ be a box in a valid box decomposition of the dense graph $\hat{G}$.
If $Q$ contains a fix subtree before Dominator's move, then Dominator gains at least $9 - k$ points in the following move, 
where $k$ is the number of points needed in order to convert the resulting graph to a good graph, and $k \leq 2$.
If $Q$ contains a strong fix subtree, then Dominator gains at least $10 - k$ points.
\end{lemma}
\Proof
Let $Q$ be a box in $\hat{G}$ as described.
If $Q$ contains a fix subtree, consider the move $v$ which is a lead of a high tail of length $2$ adjacent to $u$.
Playing $v$ converts at least $3$ vertices ($v$ and its adjacent leaf, and the $B_3$ leaf) to red, 
which gains at least $9$ points, and no new $B_2$ vertices are created.
Since $Q$ contains at most two $B_2$ vertices, at most two points 
are needed in order to convert all resulting boxes to high, and therefore $k \leq 2$.

For a strong fix subtree, if $9$ points are gained then no additional vertices became red (otherwise the gain would be greater than $9$). 
Therefore, if $u$ is converted to $B_2$, one of the following cases occurs.

\smallskip
\par\noindent
{\bf Case ($1$):}
$u$ is a strong fix vertex with a white neighbor that is the lead of a tail of length $2$ that is not white. 
Then the resulting box is a path of the form $BWB_2HH$, and it can be converted to a regular colored path box of the form $B_2WB_2HH$ (if it is not already so).

\smallskip
\par\noindent
{\bf Case ($2$):}
Either $u$ does not have additional neighbors, or its neighbors are not leads of tails of length $2$.
After converting all $B_2$ vertices except $u$ in the box containing $u$ to $B_3$, and possibly disconnecting edges between $u$ and its blue neighbors (so that $u$ has at most one neighbor), 
$u$ and the remaining subtail from the fix subtree can be separated into a dispensible box of type $1$, with $u$ as the box root.

\smallskip
\par\noindent
In both cases, at least $10 - k$ points are gained.
\QED

\begin{lemma}
\label{lemma:semi_triplet_subtree_gain}
Let $Q$ be a box in a valid box decomposition of the dense graph $\hat{G}$.
If $Q$ contains a semi-triplet subtree before Dominator's move, then the following properties hold.
\begin{enumerate}
	\dnsitem If the semi-triplet subtree has a $B_3$ leaf, then at least $11 - k$ points are gained in the following move,
	where $k$ is the number of points needed in order to convert the resulting graph to a good graph, and $k \leq 2$.
	\dnsitem If the semi-triplet subtree is strong, then at least $8 - k$ points are gained in the next move.
	\dnsitem Otherwise, at least $7 - k$ points are gained.
\end{enumerate}
\end{lemma}
\Proof
Let $Q$ be a box in $\hat{G}$ as described that contains a semi-triplet subtree rooted at a vertex $u$.
We analyze the different cases.
\smallskip
\par\noindent
{\bf Case ($1$):}
The semi-triplet subtree rooted at $u$ contains a $B_3$ leaf, $\lambda$.
If $u$ is played, then at least $3 + 3 + 3 + 1 + 1 = 11$ points are gained from the vertices that become red ($u$, $\lambda$ and the neighbor of $\lambda$) and from converting $B$ vertices in resulting components of size $2$ to $B_2$. We note that if less than two $BW$ components are created as a result of playing $u$, then at least one additional vertex was converted to red and therefore at least $3 \cdot 4 = 12$ points are gained.
The claim follows, since $Q$ contains at most two $B_2$ vertices and therefore $k \leq 2$. 

\smallskip
\par\noindent
{\bf Case ($2$):}
The semi-triplet subtree rooted at $u$ does not contain a $B_3$ leaf.
Then it must contain a $B_3$ tail lead, $u_1$. 
Let $v$ be another tail lead. As a result of playing $v$, at least $3 + 3 + 1 = 7$ points can be gained from the resulting red vertices ($v$ and the adjacent leaf), and from disconnecting $u_1$ to a component of size $2$ and converting it to $B_2$. 
As before, $k \leq 2$, and at least $7 - k$ points are gained.

If the semi-triplet subtree is strong, then an additional point can be gained by converting $u$ to $B_2$:
If $u$ does not have an additional white neighbor that is not in the triplet subtree, 
or if $u$ has such a neighbor and it is not the lead of a subtail of length $1$ or $2$, 
then $u$ can be a box root of a dispensible box of type $1$ (possibly after disconnecting edges between $u$ and its blue neighbors). 
Otherwise, $u$ has a white neighbor that is the lead of a subtail of length $2$, and this subtail is not white (note that $u$ cannot have a white leaf neighbor by the definition of strong semi-triplet). 
We conclude that $u$ is in a box of the form $BWB_2HH$, and therefore it can be converted to a regular colored path box by converting all $B$ vertices to $B_2$.
\QED

\subsection{Results of Staller moves}
\label{sub:an_staller}

In this section we analyze all possible Staller moves, i.e., the result of Staller playing any vertex $m_{2t}$ of the underlying graph.
Notice that $m_{2t}$ may be in the dense graph, or in a triplet subtree in the underlying graph.
Theorem \ref{thm:staller_good} summarizes all the possible outcomes of Staller moves.

\begin{theorem}
\label{thm:staller_good}
If Staller plays on a vertex $v$ in $G_{t-1}$ and $G_{t-1}$ is good, then at least $3$ points are gained and the resulting underlying graph, $G_t$, is good.
\end{theorem}

We separate the proof of the theorem into several claims, and note that Lemma \ref{claim:ana_box} guarantees that it suffices to analyze each move inside the box containing it (and calculate the gain on the underlying graph accordingly).

\paragraph{}

First, we extend the definition of box decomposition to the underlying graph in the following natural way:
\begin{definition}
A decomposition $\mathcal{Q}$ of the set $V_t$ of vertices of the underlying graph $G_t$ is called a \emph{box decomposition}, if the decomposition $\hat{\mathcal{Q}}_t$ which results from $\mathcal{Q}$ by replacing (without repetitions) each vertex of the underlying graph $G_t$ that is not on the dense graph $\hat{G}_t$ with the virtual leaf that replaces them on $\hat{G}_t$ (i.e., the vertex $z_{2i+1}$ adjacent to the nearest triplet head) is a box decomposition of $\hat{G}_t$.
\end{definition}

\begin{claim}
\label{claim:staller_triplet}
If Staller plays on $G_{t-1}$ a vertex $v$ that is not on the dense graph $\hat{G}_{t-1}$, then at least $3$ points are gained and the resulting dense graph $\hat{G}_t$ is good.
\end{claim}
\begin{figure}[thbp] 
  \caption{\sf All possible Staller moves on $G_{t-1}$ (marked as $v$) on triplet witnesses as described in the proof of Claim \ref{claim:staller_triplet}.
	Each tail represents the subtree of a triplet witness, which is either a real tail (of the witness is in $\mathcal{WT}_2$), or a triplet subtree. 
	In all cases, (a) is the graph $G_{t-1}$ before the move and (b) is the resulting graph $G_t$.}
  \medskip
  \centering
    \fbox{\includegraphics[width=14cm]{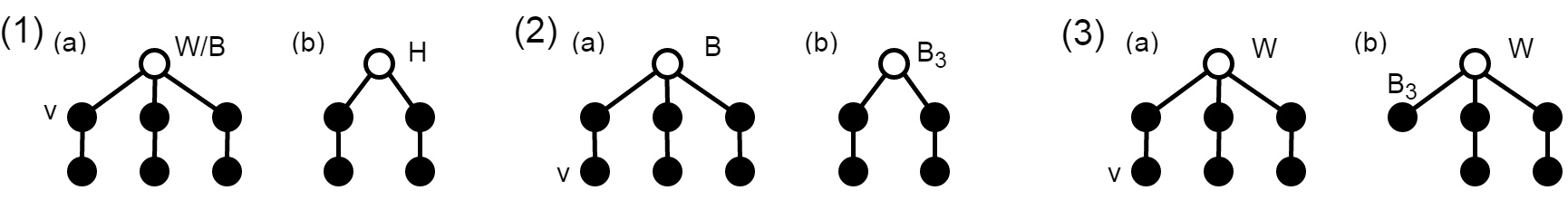}}
  \label{fig:triplet_staller}
\end{figure}
\Proof
If $v$ is not on the dense graph $\hat{G}_{t-1}$, then it is in a triplet subtree rooted at a vertex $u$, and $u$ is in some box $Q$ on $\hat{G}_{t-1}$.
There are three cases to consider, illustrated in Figure \ref{fig:triplet_staller}.
\bigskip
\par\noindent
{\bf Case ($1$):}
$v$ is not a leaf.
Then at least $3 + 3 = 6$ points can be gained from $v$ and its triplet witnesses or its adjacent leaf, 
and all resulting components on $\hat{G}_t$ are $BW$ components (if $v$ was in $\mathcal{TT}$), and the component containing $u$.
Since $Q$ contains at most two $B_2$ vertices, at least $6 - 2 = 4$ points can be gained while converting the boxes resulting from $Q$ (that are not $BW$ components) to high, 
and therefore Invariant $\invboxes$ is satisfied, so the resulting dense graph $\hat{G}_t$ is good.

\bigskip
\par\noindent
{\bf Case ($2$):}
$v$ is a leaf at distance $2$ from $u$, and $u$ is blue.
Then at least $6$ points are gained since $v$ and its neighbor become red. Therefore, as before,
at least $6 - 2 = 4$ points can be gained while satisfying Invariant $\invboxes$.

\bigskip
\par\noindent
{\bf Case ($3$):}
$v$ is a leaf and the nearest split vertex $v_1$ is white.
Therefore $3$ points are gained, and the resulting box $Q'$ contains a fix vertex ($v_1$).
If $Q'$ is corrupted, then at least one of the following three subcases occurs.

\smallskip
\par\noindent
{\bf Subcase ($3.1$):}
$v_1$ was not a triplet head in $Q$. Then $v_1$ is a strong fix vertex in $Q'$, and Lemma \ref{lemma:fix_subtree_gain} guarantees that at least $10 - 2 = 8$ 
points can be gained while converting $Q'$ to a high box, and therefore the component is semi-corrupted. Therefore Invariant $\invboxes$ is satisfied by the resulting dense graph $\hat{G}_t$.

\smallskip
\par\noindent
{\bf Subcase ($3.2$):}
$Q$ contained a single $B_2$ vertex. Then Lemma \ref{lemma:fix_subtree_gain} guarantees that at least $9 - 1 = 8$ points can be gained while converting $Q'$ to a high box, and as in the previous item, this implies that Invariant $\invboxes$ is satisfied by the resulting dense graph $\hat{G}_t$.

\smallskip
\par\noindent
{\bf Subcase ($3.3$):}
$Q$ contained two $B_2$ vertices.
This subcase splits further into three.
\begin{description}
	\dnsitem[Subcase ($3.3.1$):] If $Q$ was a dispensible box of type $2$, then $Q'$ contains a strong fix vertex 
	(since all white vertices in the dense graph had at most one white neighbor, and if this neighbor was a virtual leaf then they have no additional white neighbors). 
	From Lemma \ref{lemma:fix_subtree_gain} we conclude that $Q'$ is semi-corrupted, therefore Invariant $\invboxes$ is satisfied by the resulting dense graph $\hat{G}_t$.
	\dnsitem[Subcase ($3.3.2$):] If $Q$ was a $C_{12}$ box, then again $Q'$ contains a strong fix vertex (since all neighbors of white leaves in the dense graph, i.e., all vertices that could be triplet vertices in the underlying graph, have at most one white neighbor, and this neighbor is not the lead of a white tail of length $1$ or $2$). 
	\dnsitem[Subcase ($3.3.3$):]If $Q$ was a regular colored box with two $B_2$ vertices, then $Q'$ is not corrupted, and therefore this case can be ignored.
\end{description}
Note that the above subcases cover all graphs in which $Q'$ is corrupted, for the following reasons:
If $Q$ was a high leftover or high regular box then $Q'$ is not corrupted.
Additionally, if $Q$ was a regular colored box satisfying Property $\propbbr$ (Subcase ($3.3.3$)), then $Q'$ also satisfies it.
Finally, we know that $Q$ was not corrupted because Invariant $\invboxes$ guarantees that there are no corrupted boxes before Staller's move.
\QED

\begin{claim}
\label{claim:staller_dispensible}
If Staller plays on $\hat{G}_{t-1}$ a vertex $v$ that is in a dispensible box, then at least $3$ points are gained and the resulting dense graph $\hat{G}_t$ is good.
\end{claim}
\begin{figure}[thbp]
  \caption{\sf All possible moves on dispensible boxes.
		(a) Dispensible box of type $1$.
		(b), (c) Dispensible box of type $2$.}
  \medskip
  \centering
    \fbox{\includegraphics[width=12cm]{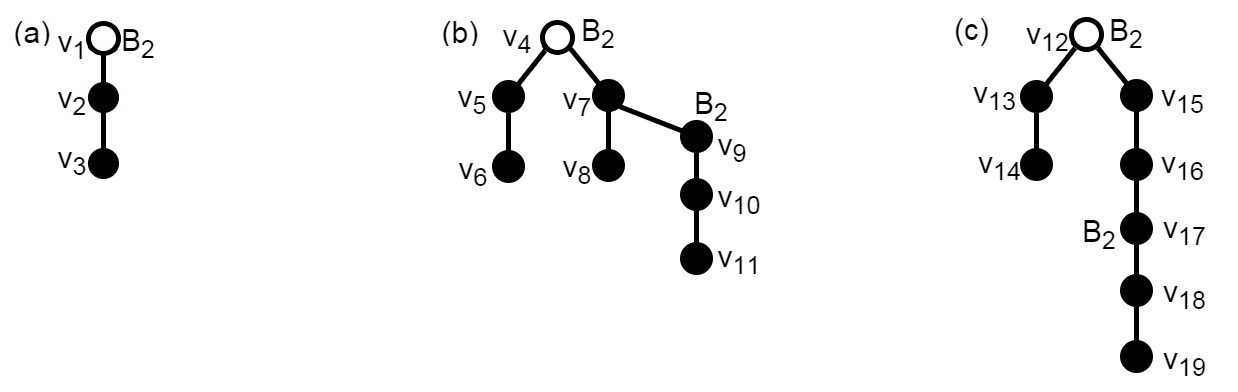}}
  \label{fig:dispensible_staller}
\end{figure}

\begin{figure}[hbtp]
  \caption{\sf Possible boxes resulting from Staller moves on dispensible boxes.}
  \medskip
  \centering
    \fbox{\includegraphics[width=7cm]
{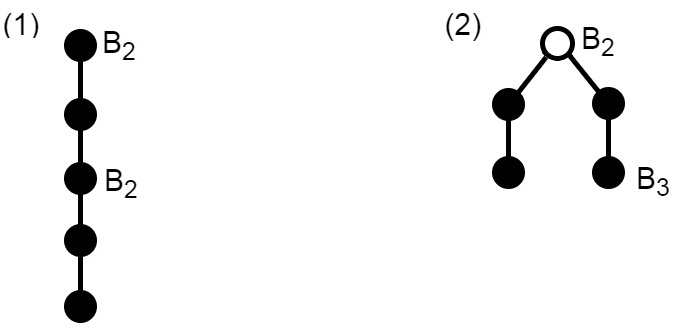}}
  \label{fig:dispensible_staller_sc}
\end{figure}
\Proof

Let $Q$ be a dispensible box.
The possible moves are shown in Figure \ref{fig:dispensible_staller}.

\smallskip
\par\noindent
{\bf Case (a):}
$Q$ is dispensible of type $1$.
Then Staller has two choices.
\begin{enumerate}
	\dnsitem If Staller plays $v_1$, then at least $3$ points are gained from $v_1$ and $v_2$, and a single $BW$ component results. From Lemma \ref{claim:ana_box} we know that Invariant $\invboxes$ is satisfied on $\hat{G}_t$. 
	\dnsitem If Staller plays $v_2$ or $v_3$, then from Lemma \ref{claim:ana_box} we know that at least $3 + 3 + 2 - 3 = 5$ points are gained and the resulting graph $\hat{G}_t$ is good.
\end{enumerate}

\smallskip
\par\noindent
{\bf Cases (b) and (c):}
$Q$ is dispensible of type $2$.
Then Staller has three types of choices.
\begin{enumerate}
	\dnsitem If Staller plays a high vertex, i.e., $v_5$, $v_6$, $v_7$, $v_8$, $v_{10}$, $v_{11}$, $v_{13}$, $v_{14}$, $v_{15}$, $v_{16}$, $v_{18}$ or $v_{19}$,
	then at least $6 - 2 = 4$ points are gained and the resulting boxes are high, and therefore Invariant $\invboxes$ is satisfied.
	\dnsitem If Staller plays $v_4$, $v_9$ or $v_{12}$, then at least $4$ points are gained, and the resulting boxes are dispensible, regular of size $2$, and if $v_{12}$ is played then regular colored path (see Case (1) in Figure \ref{fig:dispensible_staller_sc}).
	\dnsitem If Staller plays $v_{17}$, then at least $3$ points are gained and the resulting box is a path of the form $HHB_2HB_3$ (see Case (2) in Figure \ref{fig:dispensible_staller_sc}). The containing component is semi-corrupted since Dominator can play on the middle $B_2$ vertex of this box (which is the box root) and gain at least $2 + 3 + 3 + 1 = 9$ points. 
\QED
\end{enumerate}

\begin{claim}
\label{claim:staller_clean_leftover}
If Staller plays on $\hat{G}_{t-1}$ a vertex $v$ that is in a high leftover box, then at least $3$ points are gained and the resulting dense graph $\hat{G}_t$ is good.
\end{claim}
\Proof
If $v$ is in a high leftover box, then at least $3$ points are gained from $v$ becoming red, and all resulting boxes are high and do not contain triplet subtrees.
Therefore there is a box decomposition satisfying Invariant $\invboxes$ on the resulting dense graph $\hat{G}_t$.
\QED

\begin{claim}
\label{claim:staller_regular}
If Staller plays on $\hat{G}_{t-1}$ a vertex $v$ that is in a regular box, then at least $3$ points are gained and the resulting dense graph $\hat{G}_t$ is good.
\end{claim}
\Proof
Let $Q$ be a regular box.
There are four cases to consider, corresponding to the four categories in Definition \ref{def:regular_box}.

\bigskip
\par\noindent
{\bf Case $(\regtwo)$:}
$Q$ is of size $2$.
Then any move on $Q$ gains at least $5$ points, and Lemmas \ref{claim:ana_dense} and \ref{claim:ana_box} guarantee that the resulting graph is good.

\bigskip
\par\noindent
{\bf Case $(\regclean)$:}
$Q$ is a high regular box.
Then any move gains at least $3$ points and all resulting boxes can be high.

\begin{figure}[t]
  \caption{\sf All possible moves on a $C_{12}$ box. The dotted edges correspond to the two types of $C_{12}$ boxes, and exactly one of them exists.}
  \medskip
  \centering
    \fbox{\includegraphics[width=5cm]{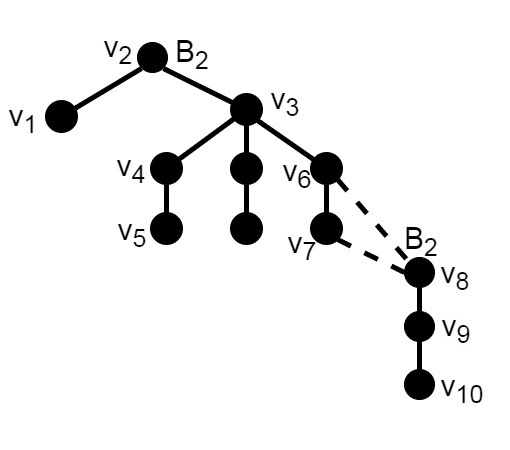}}
  \label{fig:c12_staller}
\end{figure}

\bigskip
\par\noindent
{\bf Case $(\regct)$:}
$C$ is a $C_{12}$ box.
The possible moves are described in Figure \ref{fig:c12_staller}.

\begin{figure}[thbp]
  \caption{\sf Possible semi-corrupted boxes resulting from Staller moves on $C_{12}$ boxes, and moves $v$ gaining at least $9$ points.}
  \medskip
  \centering
    \fbox{\includegraphics[width=10cm]{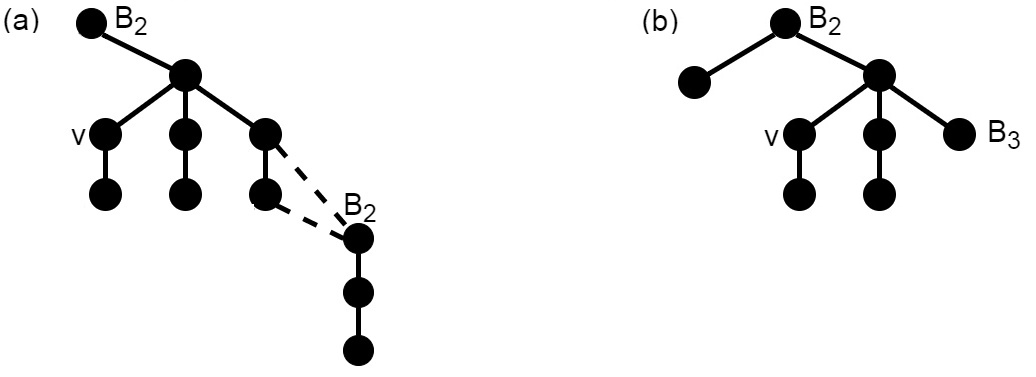}}
  \label{fig:c12_staller_sc}
\end{figure}

If Staller plays $v_1$, 
then the remaining box is semi-corrupted since Dominator can play $v_4$ in the next move and gain at least $3 + 3 + 2 + 1 = 9$ points from the red vertices and from converting $v_3$ to $B_2$, 
and the remaining box is a dispensible box of type $2$. 
See Case (a) in Figure \ref{fig:c12_staller_sc}.

If Staller plays $v_2$, then at least $5$ points are gained and, after separating the vertices $v_8$, $v_9$ and $v_{10}$ to a dispensible box of type $1$ rooted at $v_8$,
the resulting box is high and satisfies Property $\propt1$, and therefore it is regular.

If Staller plays $v_3$, then at least $6$ points are gained and all resulting boxes are regular colored path boxes and regular boxes of size $2$.

If Staller plays one of the vertices $v_4$, $v_6$, $v_9$ or $v_{10}$, then at least $6 - 2 = 4$ points can be gained while converting the $B_2$ vertices to high, and the resulting boxes are regular.

If Staller plays $v_5$ or $v_7$, then at least $3$ points are gained and the resulting box (after separating $v_8$, $v_9$ and $v_{10}$ to a dispensible box of type $1$ rooted at $v_8$)
contains a strong fix vertex and a single $B_2$ vertex (see Case (b) in Figure \ref{fig:c12_staller_sc}), and therefore it is in a semi-corrupted component. 

If Staller plays $v_8$, then at least $3$ points are gained from converting $v_9$ to a $B_2$ vertex in a $BW$ box, and $v_3$ becomes a strong semi-triplet vertex in a corrupted box.
From Lemma \ref{lemma:semi_triplet_subtree_gain} we conclude that the box is semi-corrupted, since it is possible to gain at least $8$ points in the following move (since $k = 0$).

\bigskip
\par\noindent
{\bf Case $(\regcolored)$:}
$Q$ is a regular colored box.
Case $\propbr$:(\ref{invnp:d1}) (i.e., $Q$ is a dispensible box of type $1$) was already analyzed in Claim \ref{claim:staller_dispensible}, and therefore we ignore it.
Therefore exactly one of the following cases occurs (see Figure \ref{fig:regular_complex}).
Note that in all the following cases, we analyze the subtree rooted at some split vertex, and show that all the resulting boxes are regular. 
If an unexpected vertex becomes red, then at least two additional points are gained, and therefore the box containing this vertex can be converted to a high box. 
We therefore ignore this possibility in the case analysis.
We split the analysis into two subcases, as follows.

\smallskip
\par\noindent
{\bf Subcase $(a)$:}
$v$ is a $B_2$ vertex. Then at least one of the following cases occurs.
	\begin{enumerate}
		\dnsitem $v$ is the only $B_2$ vertex and $Q$ satisfies Case $\propbr$:(\ref{invnp:1leaf}) of Definition \ref{def:regular_colored}. 
		Then at least $2 + 1 = 3$ points are gained by converting the vertex $v'$ adjacent to $v$ to $B_2$. 
		If the resulting box does not satisfy Case $\propbr$:(\ref{invnp:1leaf}), then it must satisfy Case $\propbr$:(\ref{invnp:1tail}), and therefore it is a regular colored box.
		
		\dnsitem $v$ is the only $B_2$ vertex and $Q$ satisfies Case $\propbr$:(\ref{invnp:1tail}).
		Then at least $2 + 1 = 3$ points are gained from $v$ and from an adjacent subtail lead, and the resulting boxes are a high box, 
		and regular colored path boxes satisfying case $\propbr$:(\ref{invnp:1tail}) or $\propbr$:(\ref{invnp:d1}) 
		(and possibly additional $BW$ components). 
		
		\dnsitem $v$ is a $B_2$ leaf and $Q$ satisfies Case $\propbbr$:(\ref{invnp:2leaves}) or Case $\propbbr$:(\ref{invnp:2tail}). 
		Then, similarly to Case $\propbr$:(\ref{invnp:1leaf}), at least $3$ points are gained from $v$ and its neighbor, and if the resulting box cannot be converted to a high box while gaining at least $3$ points, 
		then it satisfies one of the cases $\propbbr$:(\ref{invnp:2leaves}) and $\propbbr$:(\ref{invnp:2tail}).
		
		\dnsitem $v$ is a non-leaf vertex and $Q$ satisfies Case $\propbbr$:(\ref{invnp:2tail}).
		Then, similarly to Case $\propbr$:(\ref{invnp:1tail}), at least $3$ points are gained from $v$ and from an adjacent tail lead, and the resulting boxes are a high box, and regular colored path boxes 
		(and possibly additional $BW$ components). 
	\end{enumerate}

\smallskip
\par\noindent
{\bf Subcase $(b)$:}
$v$ is a high vertex. Then at least $3$ points are gained from $v$, and therefore no additional $B_2$ vertices need to be created. At least one of the following cases occurs.
	\begin{enumerate}
		\dnsitem $Q$ contains a single $B_2$ vertex, $v'$, satisfying Case $\propbr$:(\ref{invnp:1leaf}) of Definition \ref{def:regular_colored}.
		Let $u$ be the vertex closest to $v'$ that has a subtail ending at $v'$ and another subtail (guaranteed to exist by the definition).
		Exactly one of the following cases occurs.
		\begin{enumerate}
			\dnsitem $v$ is on the subtail of $u$ that contains $v'$, or $v = u$.
			Then at least $3 + 1 = 4$ points are gained, 
			and the resulting boxes are high regular boxes, and possibly a regular colored path box with two $B_2$ leaves.
			\dnsitem $v$ is on another subtail of $u$.
			Then at least $3 + 1 = 4$ points can be gained, 
			and the resulting boxes are a regular colored box satisfying Case $\propbbr$:(\ref{invnp:2leaves}) or Case $\propbbr$:(\ref{invnp:2tail}), and possibly another regular colored path with a single $B_2$ leaf.
			\dnsitem $v$ is not on a subtail of $u$.
			Then at least $3$ points are gained and the resulting box that is not high can satisfy Case $\propbr$:(\ref{invnp:1leaf}).
		\end{enumerate}
		
		\dnsitem $Q$ contains a single $B_2$ vertex, $u$, satisfying Case $\propbr$:(\ref{invnp:1tail}). Then exactly one of the following cases occurs.
		Notice that $u$ is blue and therefore does not have neighbors that are blue leaves, which implies that it does not have leaf neighbors.
		\begin{enumerate}
			\dnsitem $v$ is on a subtail of $u$, and is at distance $1$ or $2$ from $u$. 
			Then at least $4$ points can be gained from $v$ and its neighbors, which means that at least $3$ points can be gained while converting the box containing $u$ to a high box.
			\dnsitem $v$ is on a subtail of $u$, and is at distance $3$ or more from $u$.
			Then at least $4$ points can be gained, and the resulting boxes are a regular colored box satisfying Case $\propbbr$:(\ref{invnp:2tail}), and possibly a regular colored path with a single $B_2$ leaf, or a $BW$ component.
			\dnsitem $v$ is not on a subtail of $u$. Then at least $3$ points can be gained, and the resulting boxes can be high boxes and a regular colored box satisfying Case $\propbr$:(\ref{invnp:1tail}).
		\end{enumerate}
		
		\dnsitem $Q$ contains two $B_2$ vertices, $v_1$ and $v_2$, that are leaves of subtails of a vertex $u$, corresponding to Case $\propbbr$:(\ref{invnp:2leaves}).
		Then exactly one of the following cases occurs.
		\begin{enumerate}
			\dnsitem $v$ is on a subtail of $u$ that contains a $B_2$ vertex, or $v = u$. Then at least $4$ points can be gained on the subtail, 
			and at least $3$ points can be gained while converting the other $B_2$ leaf to $B_3$.
			In this case, all resulting boxes are high boxes and possibly a regular colored path with two $B_2$ leaves.
			\dnsitem Otherwise, at least $3$ points are gained and the resulting box that is not high still satisfies Case $\propbbr$:(\ref{invnp:2leaves}).
		\end{enumerate}
		
		\dnsitem $Q$ contains two $B_2$ vertices, $v_1$ and $v_2$, such that $v_1$ is a leaf on a subtail of $v_2$, corresponding to Case $\propbbr$:(\ref{invnp:2tail}).
		Then exactly one of the following cases occurs.
		\begin{enumerate}
			\dnsitem $v$ is on the subtail of $v_2$ that contains $v_1$. Then at least $4$ points can be gained on the subtail, 
			which means that at least $3$ points can be gained while converting $v_2$ to $B_3$.
			In this case, all resulting boxes are high, and possibly a regular colored path with two $B_2$ leaves.
			\dnsitem Otherwise, at least $3$ points are gained and the resulting box that is not high still satisfies Case $\propbbr$:(\ref{invnp:2leaves}).
\QED
		\end{enumerate}
	\end{enumerate}

Theorem \ref{thm:staller_good} follows from Claims \ref{claim:staller_triplet}, \ref{claim:staller_dispensible}, \ref{claim:staller_clean_leftover} and \ref{claim:staller_regular}.

\subsection{Dominator moves}
\label{sub:an_dom}

We have seen that if Staller plays on a good graph $G_t$, 
then the resulting graph is also good.
Our goal in this section is to prove the following theorem.

\begin{theorem}
\label{thm:dom_gain}
Let $t < T$.
If Dominator plays on a vertex $v$ in $\hat{G}_{t-1}$ and $\hat{G}_{t-1}$ is good, and Dominator chooses all moves greedily according to the guidelines in Section \ref{section:algorithm_outline},
then the resulting graph $\hat{G}_t$ is good, and at least one of the following properties holds.
\begin{enumerate}
	\dnsitem $\psi_t \geq 0$.
	\dnsitem $\psi_t \geq -2$ and $\psi_t + \psi_{t+1} \geq 0$.
	\dnsitem $\psi_t \geq -2$ and $\psi_t + \psi_{t + 1} + \psi_{t + 2} \geq 0$.
\end{enumerate}
\end{theorem}

Notice that we do not make requirements about the last move (step $T$) because of Corollary \ref{cor:last_5}, and therefore in all the following claims we only consider $t$ such that $t < T$.

The definition of semi-corrupted components guarantees that if a semi-corrupted component is created, then in the following Dominator move $\psi \geq 1$, and therefore we focus on the case that there is a box decomposition that does not contain corrupted boxes.
Let $\mathcal{Q}$ be a box decomposition of $\hat{G}_{t-1}$ that does not contain corrupted boxes. 

\begin{claim}
\label{claim:c12_dom}
If $\mathcal{Q}$ contains a $C_{12}$ box, then $\psi_t \geq 0$ and the resulting graph $\hat{G}_t$ is good.
\end{claim}
\Proof
Recall that all possible moves appear in Figure \ref{fig:c12_staller}. 
If Dominator plays $v_4$, then at least $3 + 3 + 1 = 7$ points are gained and the resulting boxes, after disconnecting the edge between $v_2$ and $v_3$, are a $BW$ box and a dispensible box of type $2$.
\QED

\begin{claim}
\label{claim:b2_dom}
If $\mathcal{Q}$ contains a regular colored box (including a dispensible component of type $1$),
then $\psi_t \geq 0$ and the resulting graph $\hat{G}_t$ is good.
\end{claim}
\begin{figure}[thbp]
  \caption{\sf Possible moves (marked as $v$) gaining at least $7$ points on regular colored boxes containing $B_2$ vertices, corresponding to the different cases in the proof of Claim \ref{claim:b2_dom}.
		(a) Case ($\propbr$:\ref{invnp:1leaf}). 
		(b) Case ($\propbr$:\ref{invnp:1tail}). 
		(c) Case ($\propbr$:\ref{invnp:d1}).
		(d) Case ($\propbbr$:$1$). 
		(e) Case ($\propbbr$:$2$).}
  \medskip
  \centering
    \fbox{\includegraphics[width=14cm]{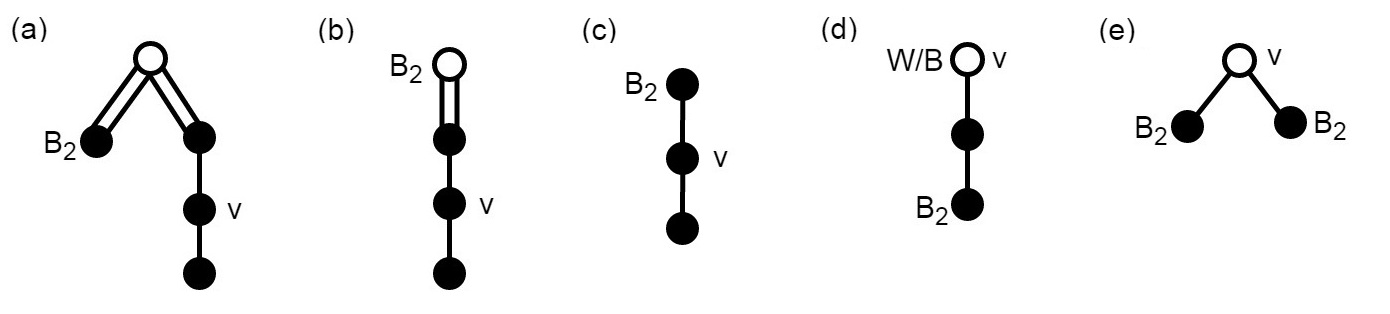}}
  \label{fig:regular_colored_b2_dom}
\end{figure}
\Proof
Let $Q$ be a regular colored box.
The following cases cover all possibilities for $Q$.

Except where noted otherwise, we assume all high vertices are white. If an unexpected vertex becomes red, then at least two additional points are gained, and the resulting box can be converted to a high box. Therefore we ignore this possibility in the analysis.

See Figure \ref{fig:regular_colored_b2_dom} for illustrations, and recall that all possible regular colored boxes are illustrated in Figure \ref{fig:regular_complex}.
We separate the analysis into cases according to the different properties of Definition \ref{def:regular_colored}, as follows.
\bigskip
\par\noindent
{\bf Case ($\propbr$:\ref{invnp:1leaf}):}
$Q$ satisfies Case $\propbr$:(\ref{invnp:1leaf}) of Definition \ref{def:regular_colored}.
Let $u$ be a vertex with a subtail containing a $B_2$ leaf and a high subtail of length $3$ or more.
By playing on the neighbor of the leaf on the high subtail,
at least $3 + 3 + 1 = 7$ points can be gained and the resulting box satisfies Case $\propbbr$:(\ref{invnp:2leaves}).

\bigskip
\par\noindent
{\bf Case ($\propbr$:\ref{invnp:1tail}):}
$Q$ satisfies Case $\propbr$:(\ref{invnp:1tail}). 
Let $u$ be the $B_2$ vertex. 
By playing on the neighbor of the leaf on a subtail of $u$ of length $3$ or more, exactly one of the following cases can result.
\begin{enumerate}
	\dnsitem At least $3 + 3 + 1 = 7$ points are gained, and the resulting box satisfies Case $\propbbr$:(\ref{invnp:2tail}).
	\dnsitem At least $3 + 3 + 3 - 1 = 8$ points are gained after converting $u$ to $B_3$, and the resulting box is high.
\end{enumerate}

\smallskip
\par\noindent
{\bf Case ($\propbr$:\ref{invnp:d1}):}
$Q$ satisfies Case $\propbr$:(\ref{invnp:d1}). 
By playing on the middle vertex $v$, 
all vertices of the box are eliminated and at least $2 + 3 + 3 = 8$ points are gained.

\bigskip
\par\noindent
{\bf Case ($\propbbr$):}
$Q$ satisfies Case $\propbbr$:(\ref{invnp:2leaves}) or Case $\propbbr$:(\ref{invnp:2tail}).
This splits further into the following two subcases.

\smallskip
\par\noindent
{\bf Subcase ($1$):}
	There is a $B_2$ leaf on a subtail of length $2$ of some vertex $v$. Since the leaf is a $B$ vertex, its neighbor must be white.
	Exactly one of the following cases results from playing $v$.
	\begin{enumerate}
		\dnsitem $v$ is high. Then at least $2 + 3 + 3 - 1 = 7$ points can be gained while converting the remaining $B_2$ vertex in the box to high.
		\dnsitem $v$ is not high. Then at least $2 +3 + 2 = 7$ points are gained and the resulting boxes do not contain $B_2$ vertices.
	\end{enumerate}

\par\noindent
{\bf Subcase ($2$):}
	Otherwise, there must be two $B_2$ leaves that are neighbors of the same vertex, $v$.
	Playing $v$ gains at least $2 + 3 + 2 = 7$ points, and the resulting boxes are high.
\QED

\begin{claim}
\label{claim:dom_split_7}
If $\mathcal{Q}$ contains a high regular box $Q$ of size $3$ or more (including a high leftover root box), and $Q$ is either a path or contains a split vertex $s$ satisfying one of the following requirements:
\begin{enumerate}
	\dnsitem 
	\label{claim:dom_split_7:4tails}
	$s$ has four tails or more.
	\dnsitem 
	\label{claim:dom_split_7:not2}
	$s$ has a tail that is not of length $2$.
	\dnsitem
	\label{claim:dom_split_7:b3}
	$s$ has a tail of length $2$ containing a $B_3$ vertex.
\end{enumerate}
Then $\psi_t \geq 0$ and the resulting graph $\hat{G}_t$ is good.
\end{claim}
\begin{figure}[thbp]
  \caption{\sf Possible moves (marked as $v$) gaining at least $7$ points on high regular complex boxes, corresponding to the different cases in the proof of Claim \ref{claim:dom_split_7}.
		When there are two items, (a) is before playing $v$ and (b) is the result.
		(1) Case (a).
		(2) Case (b). 
		(3) Case (c:1). 
		(4) Case (c:2). 
		(5) Case (d:1). 
		(6) Case (d:2).} 
  \medskip
  \centering
    \fbox{\includegraphics[width=11cm]{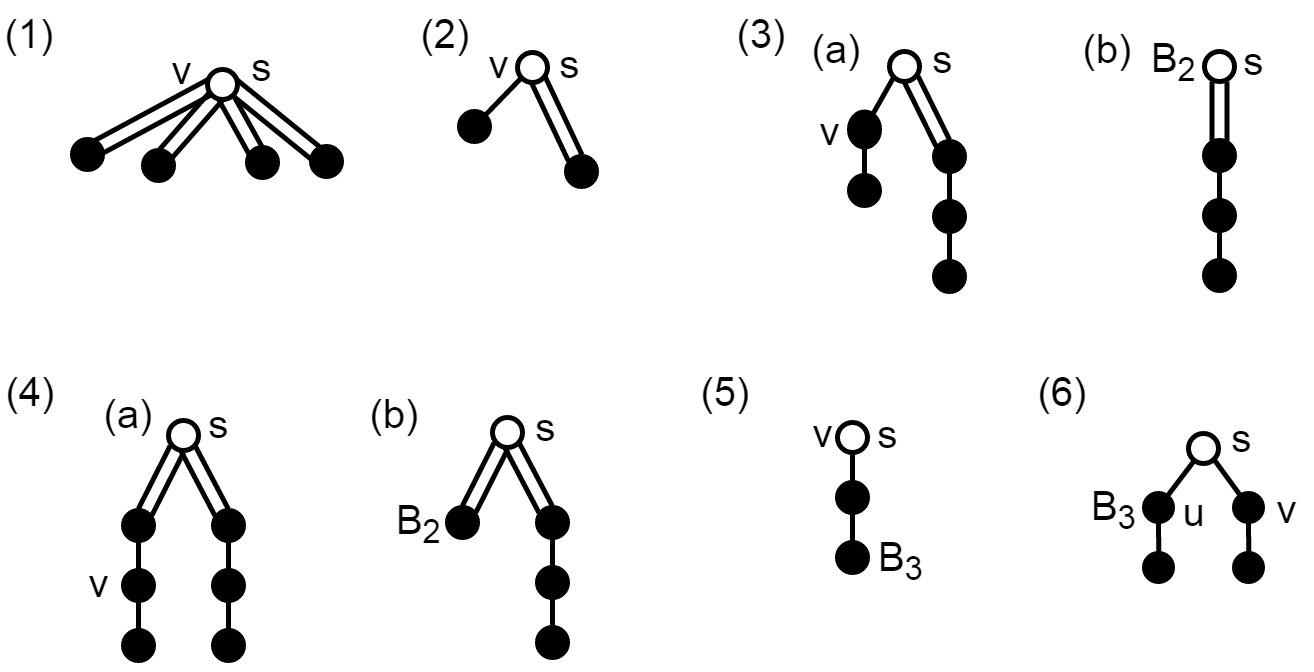}}
  \label{fig:dom_clean}
\end{figure}
\Proof
First, observe that if $\mathcal{Q}$ contains a high path of length $3$ or more, then playing on the neighbor of a leaf on this path can gain at least $3 + 3 + 1 = 7$ points, 
and if a box remains, it is a path with a $B_2$ leaf and no other $B_2$ vertices, and therefore it is a regular box. 
Otherwise, let $Q$ be a high regular complex box, and $s$ a split vertex in $Q$, as described.
As before, we assume all high vertices are white, since if an unexpected vertex becomes red, at least two additional points are gained and the box containing the red vertex can be converted to a high box.
We separate the analysis into cases according to the different conditions of the claim.
See Figure \ref{fig:dom_clean} for illustrations.

\bigskip
\par\noindent
{\bf Case (a):}
$s$ has $4$ tails or more.
Then playing $s$ gains at least $3 + 4 = 7$ points, and the resulting boxes are high boxes, and paths with a $B_2$ leaf and no internal $B_2$ vertices, i.e., regular colored path boxes and boxes of size $2$.

\smallskip
\par\noindent
{\bf Case (b):}
$s$ has a leaf neighbor and an additional tail.
Then playing $s$ can gain at least $3 + 3 + 1 = 7$ points from the red vertices and the other tail lead, and as in the previous case, the resulting boxes are high boxes, 
regular colored path boxes and boxes of size $2$.

\smallskip
\par\noindent
If Cases (a) and (b) are not satisfied, then all tails are of length $2$ or more.

\smallskip
\par\noindent
{\bf Case (c):}
$s$ has a tail of length $3$ or more.
If Dominator plays the vertex $v$ that is the neighbor of the leaf on the shortest tail, then exactly one of the following two subcases occurs.
\smallskip
\par\noindent
{\bf Subcase (1):}
	The shortest tail is of length $2$. 
	Then playing $v$ gains $3 + 3 + 1 = 7$ points from the red vertices and from $s$, and the resulting box satisfies Case $\propbr$:(\ref{invnp:1tail}) of Definition \ref{def:regular_colored}.
\smallskip
\par\noindent
{\bf Subcase (2):}
	The shortest tail is of length $3$ or more.
	Then playing $v$ gains at least $3 + 3 + 1 = 7$ points from $v$ and its neighbors, and the resulting box satisfies Case $\propbr$:(\ref{invnp:1leaf}) of Definition \ref{def:regular_colored}.

\bigskip
\par\noindent
{\bf Case (d):}
All tails are of length exactly $2$, and $s$ has a tail of length $2$ containing a $B_3$ vertex.
Then at least one of the following two subcases occurs.
\smallskip
\par\noindent
{\bf Subcase (1):}
	There is a $B_3$ leaf on a tail of $s$.
	Then playing $s$ gains at least $3 + 3 + 3 + 1 = 10$ points from the vertices that become red and the tail leads, and the resulting boxes are high boxes and boxes of size $2$.
\smallskip
\par\noindent
{\bf Subcase (2):}
	$s$ has a $B_3$ tail lead, $u$.
	Then playing a vertex $v$ that is another tail lead adjacent to $s$ can gain at least $3 + 3 + 1 = 7$ points from the red vertices and from $u$, 
	and the resulting boxes are high boxes and boxes of size $2$.
\QED

\begin{claim}
\label{claim:t4_dom}
If $\mathcal{Q}$ contains a dispensible component and $\psi_t < 0$, then $\psi_t = -1$, and additionally, $\psi_{t+1} + \psi_{t+2} \geq 1$ and all resulting graphs are good.
\end{claim}
\Proof
First observe that if there is a dispensible box of type $1$ that is a root box, then $\psi_t \geq 0$ by Claim \ref{claim:b2_dom} (in fact, $\psi_t \geq 1$).
Therefore the dispensible component is of type $2$.
Recall that all possible moves appear in Figure \ref{fig:dispensible_staller}.
If Dominator plays $v_7$ or $v_{15}$ (according to the type of $D_2$ box), 
then two of the resulting components are dispensible components of type $1$ in some valid box decomposition $\mathcal{Q}_1$ of $\hat{G}_t$.
Therefore, one of the following cases occurs.

\smallskip
\par\noindent
{\bf Case (a):}
After Staller's move a semi-corrupted box is created.
Then $\psi_{t+1} + \psi_{t+2} \geq 1$.

\smallskip
\par\noindent
{\bf Case (b):}
Staller does not create a semi-corrupted box.
Then at least one dispensible component of type $1$ remains, 
and Dominator can play move $m_{t+2}$ on a high vertex in a dispensible component of type $1$ and gain at least $8$ points.

\smallskip
\par\noindent
Since Theorem \ref{thm:staller_good} guarantees that $\psi_{t+1} \geq 0$, we get the claim.
\QED

\begin{claim}
\label{claim:minus_dom}
If $\mathcal{Q}$ contains a high regular box of size $3$ or more and $\psi_t < 0$, then $\psi_t = -1$, and additionally, $\psi_{t+1} + \psi_{t+2} \geq 1$ and all resulting graphs are good.
\end{claim}
\begin{figure}[thbp]
  \caption{\sf Subtrees in $\hat{G}_{t-1}$ corresponding to the different cases in the proof of Claim \ref{claim:minus_dom}.
		(1) Case (a). 
		(2) Case (b:1). 
		(3) Case (b:2).} 
  \medskip
  \centering
    \fbox{\includegraphics[width=14cm]{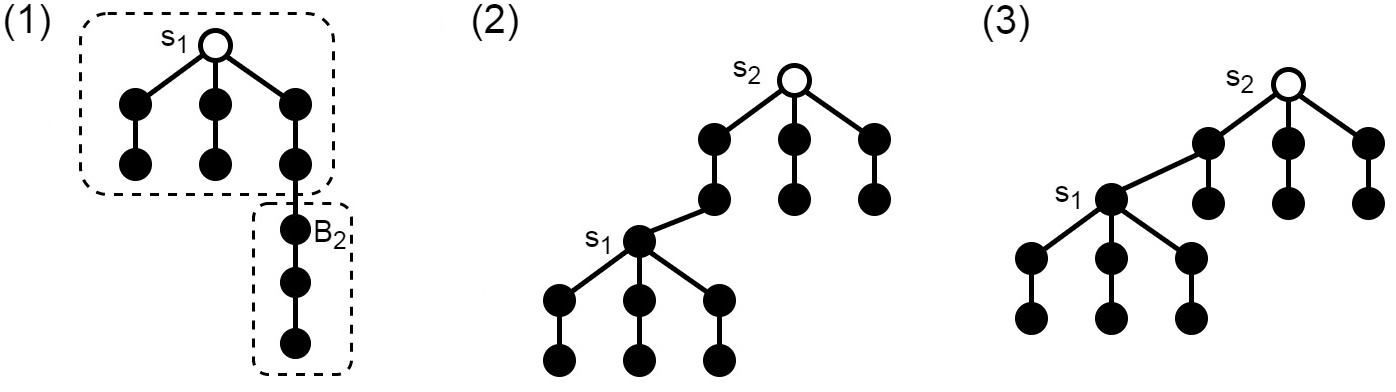}}
  \label{fig:dom_6}
\end{figure}
\Proof
From Claims \ref{claim:c12_dom}, \ref{claim:b2_dom} and \ref{claim:dom_split_7} we conclude that all regular boxes of size $3$ or more are high complex boxes, all split vertices in these boxes have two or three tails of length $2$, and all vertices in these tails are white.

Let $Q$ be a high regular box of size $3$ or more as described.
Let $s$ be a split vertex in $Q$ that has at most one neighbor that is not a tail lead (from Claim \ref{claim:every_tree_has_split} we know that such a vertex exists), and such that all vertices in the tails of $s$ are white, and assume that $\psi_t < 0$.
See illustrations in Figure \ref{fig:dom_6}.

First, observe that if there is such a split vertex $s$ with exactly two tails, then playing one of the tail leads gains at least $3 + 3 + 1 = 7$ points and the resulting box can be separated into a dispensible box of type $1$ rooted at $s$, and a high box.
Therefore, $s$ is a triplet vertex of depth $2$.
Additionally, note that Property $\propt1$ guarantees that there is a vertex $u$ on a tail of $s$ that is the parent of a dispensible box of type $1$, since if all three tail leads were potential triplet witnesses then this would imply that $s$ has a (virtual) white leaf, contradicting our assumption that all tails of $s$ are of length $2$ 
(and since a high box cannot be a parent of a high leftover box, and $\mathcal{Q}$ does not contain corrupted boxes). 
Let $v$ be a tail lead adjacent to $s$ on another tail. Playing $v$
could gain $3 + 3 + 1 = 7$ points from converting $s$ to a $B_2$ vertex that is the box root of a dispensible box of type $2$.
Since the other resulting box would be high, and $\psi_t < 0$, we conclude that doing so would violate Property $\propt1$.
Therefore, we conclude that for each such split vertex $s$ there exists a split vertex $s'$ that would become a triplet vertex in this case. 
See Cases (2) and (3) in Figure \ref{fig:dom_6} ($s_1$ corresponds to $s$ and $s_2$ corresponds to $s'$).

Let $\lambda$ be a leaf on $Q$, 
and let $s_1$ to be the split vertex farthest from $\lambda$.
Let $s_2$ be the split vertex that is closest to $s_1$.
We analyze the results of playing $m_t = s_1$, and separate them into cases according to the structure of the graph.
Observe that at least $6$ points are gained by this move, therefore $\psi_t \geq -1$.

\bigskip
\par\noindent
{\bf Case (a):}
There is a dispensible box of type $1$ that is adjacent to a leaf of a tail of $s_1$.
Then playing $s_1$ gains at least $6$ points, and the resulting boxes are $BW$ boxes, a high box containing the semi-triplet vertex $s_2$, and a path $P = (v_1, v_2, v_3, v_4, v_5)$ of the form $B_2WB_2HH$. 
See Case (1) in Figure \ref{fig:dom_6}.
For all $1 \leq i \leq 5$, 
if Staller plays $v_i$, then at least $5$ points are gained 
and in the following Dominator move there is a valid box decomposition containing a high box with a semi-triplet vertex, therefore in this case $\psi_{t+1} + \psi_{t+2} \geq 2$ by Lemma \ref{lemma:semi_triplet_subtree_gain}. 
If Staller plays elsewhere, 
then either a semi-corrupted component is created in $\hat{G}_{t+1}$ (in which case at least $8$ points are gained in step $t+2$), 
or Dominator can play $v_3$ and gain at least $2 + 3 + 2 + 1 = 8$ points in step $t+2$.
We conclude that in this case $\psi_{t+1} + \psi_{t+2} \geq 1$. 	
\bigskip
\par\noindent
{\bf Case (b):}
After Dominator plays $s_1$, the resulting graph $\hat{G}_{t+1}$ contains 
a dispensible component of type $1$ and a high box with a semi-triplet subtree rooted at $s_2$. This splits further into the following two subcases.

\smallskip
\par\noindent
{\bf Subcase (1):}
The resulting semi-triplet subtree has a $B_3$ leaf (Case (2) in Figure \ref{fig:dom_6}).
Lemma \ref{lemma:semi_triplet_subtree_gain} guarantees that in this case, if Staller does not play on the semi-triplet subtree then at least $11$ points are gained.
If Staller does play on the semi-triplet subtree, then Dominator can play on the $D_1$ component and gain at least $8$ points.
If Staller creates a semi-corrupted component, then at least $8$ points are gained in step $t+2$ as well.
In all cases, ${\psi_{t+1} + \psi_{t+2} \geq 1}$.

\smallskip
\par\noindent
{\bf Subcase (2):}
The semi-triplet subtree rooted at $s_2$ has a $B_3$ tail lead (Case (3) in Figure \ref{fig:dom_6}).
We conclude that the internal degree of $s_2$ in $\hat{G}_{t-1}$ is exactly $4$, for the following reasons:
First, assume towards contradiction that the internal degree of $s_2$ is $3$. Then $s_2$ is a split vertex with two tails of length $2$, in contradiction to the assumption that $\psi_t < 0$.
Next, assume towards contradiction that the internal degree of $s_2$ is $5$ or more.
Then $s_2$ has at least two additional neighbors, besides the tail lead $v_1$ adjacent to $s_1$ and the two other tail leads.
Since $s_1$ was chosen to be the split vertex farthest from $\lambda$, and all tails of all split vertices are of length exactly $2$, at least one of the other neighbors of $s_2$, $v_0$, must be one of the following:
\begin{enumerate}
	\dnsitem A lead of a white tail of length $2$. 
	\dnsitem A vertex of internal degree $3$ that has a white leaf and a neighbor $s_0$ that is a triplet vertex of depth $2$.
\end{enumerate}
In both cases, Dominator could play $m_t = s_2$ and gain at least $3 + 4 \cdot 1 = 7$ points from $s_2$, the tail leads and $v_1$ and $v_0$, 
and the resulting boxes in $\hat{G}_t$ would be $BW$ boxes and $C_{12}$ boxes (since every triplet subtree has a vertex that is the parent of a $D_1$ box by Property $\propt1$, as none of the split vertices have leaf neighbors).
This contradicts the assumption that $\psi_t < 0$, and we conclude that the internal degree of $s_2$ is less than $5$.

Since the internal degree of $s_2$ is exactly $4$, and it does not have another white tail of length $2$ 
or a white tail of length $1$, 
we conclude that
$s_2$ is a strong semi-triplet vertex. From Lemma \ref{lemma:semi_triplet_subtree_gain} we conclude that if Staller does not play $m_{t+1}$ on the semi-triplet subtree, then $\psi_{t+2} \geq 1$.
If Staller does play on the semi-triplet subtree, then as before, Dominator can play on the $D_1$ component and gain at least $8$ points, so either way $\psi_{t+2} \geq 1$.

\smallskip
\par\noindent
	If Staller plays elsewhere and creates a semi-corrupted component, then $\psi_{t+2} \geq 1$ as well.
\bigskip
\par\noindent
This concludes the proof, since all resulting graphs are good.
\QED

\begin{claim}
\label{claim:t1_dom}
If all root boxes in $\mathcal{Q}$ are of size $2$, then $\psi_t \geq -2$ and the resulting graph $\hat{G}_t$ is good, and at least one of the following properties holds.
\begin{enumerate}
	\dnsitem $\psi_{t+1} \geq 2$.
	\dnsitem $\psi_{t+1} + \psi_{t+2} \geq 2$.
\end{enumerate}
\end{claim}
\Proof
Recall that boxes of size $2$ cannot be parent boxes, and therefore all components of the dense graph are of size $2$ (i.e., components of the forms $B_2W$, $B_3W$ and $WW$).
Dominator can play on any real vertex and gain at least $2 + 3 = 5$ points, and the resulting dense graph $\hat{G}_t$ contains only components of size $2$.
If Staller plays on a real vertex of the dense graph, then at least $5$ points are gained and therefore $\psi_{t+1} \geq 2$.
Otherwise, since the box containing Staller's move contains at most one $B_2$ vertex, the proof of Claim \ref{claim:staller_triplet} guarantees that one of the following cases occurs.
\smallskip
\par\noindent
{\bf Case (a):}
At least $6 - 1 = 5$ points are gained in Staller's move, and the resulting box is high.
Therefore $\psi_{t+1} \geq 2$, and the resulting graph $\hat{G}_{t+1}$ is good.
\smallskip
\par\noindent
{\bf Case (b):}
At least $3$ points are gained in Staller's move (i.e., $\psi_{t+1} \geq 0$), and the resulting box is semi-corrupted and contains a strong fix vertex. 
Lemma \ref{lemma:fix_subtree_gain} guarantees that in this case, at least $10 - 1 = 9$ points are gained in the following Dominator move (i.e., $\psi_{t+2} \geq 2$), and the resulting graph $\hat{G}_{t+2}$ is good.
\QED

Theorem \ref{thm:dom_gain} follows from Claims 
\ref{claim:c12_dom}, \ref{claim:b2_dom}, \ref{claim:t4_dom}, \ref{claim:minus_dom} and \ref{claim:t1_dom}.

\subsection{Analysis conclusion}
\label{sub:an_conclusion}

We conclude by showing that if Dominator plays according to the algorithm, then the game ends with an average gain of at least $5$ points per move, and therefore Dominator wins.

\begin{theorem}
\label{cor:dom_wins}
If Dominator plays greedily according to the guidelines in Section \ref{section:algorithm_outline}, then the average gain in a Dominator-start game is at least $5$ points.
\end{theorem}
\Proof
We first note that if $\hat{G}_{t^*}$ is good for some $t^*$ and $\varexcess_{t^*} \geq 0$, then there exists $t > t^*$ such that at least one of the following properties holds.
We observe that for even $t^*$, these properties are guaranteed by Theorem \ref{thm:dom_gain} and Corollary \ref{cor:last_5},
and for odd $t^*$, they are guaranteed by Theorem \ref{thm:staller_good} (and the definition of semi-corrupted components).

\begin{enumerate}
	\dnsitem \label{cor:dom_wins:odd}
	$t$ is odd and at least $5 t + 2$ points are gained in steps $1$ through $t$, and $\hat{G}_t$ is good.
	
	\dnsitem \label{cor:dom_wins:even}
	$t$ is even and at least $5 t$ points are gained in steps $1$ through $t$, and $\hat{G}_t$ is good.
	
	\dnsitem \label{cor:dom_wins:last}
	At least $5 t$ points are gained in steps $1$ through $t$ and $\hat{G}_t$ is empty, i.e., the game is over.
\end{enumerate}

We note that for $t^* = 0$, $\hat{G}_{t^*}$ is high and therefore good,
and $\varexcess_0 = 0$, and therefore there exists some $t > 0$ satisfying one of the above cases.

The theorem follows by induction, since $\varexcess_t \geq 0$ in Cases \ref{cor:dom_wins:odd} and \ref{cor:dom_wins:even}, and the game ends when the graph is empty, and therefore $t = T$ must satisfy Case \ref{cor:dom_wins:last}.
\QED

This concludes the analysis.

\section{Implementing the algorithm}
\label{section:implementation}

The greedy algorithm described in Section \ref{section:algorithm_outline} often achieves stronger results than what is required in order to prove Conjecture \ref{conjecture:3_5}.
Specifically, it would suffice if Dominator's move was chosen such that $\varexcess$ is non-negative when possible while preserving the invariant, 
and when no such move is possible, chose a move which guarantees that the excess gain at the end of the next Staller move or the next Dominator move is non-negative (if the current Dominator move is not the last move of the game).
The analysis shown in the previous section guarantees that Dominator always has such a move.

We have implemented a variant of the algorithm in order to verify the correctness of the algorithm and the analysis, 
and ran it successfully on all trees up to size $20$ (using the tree generation algorithm described in \cite{wright1986constant}),
as well as on some specifically constructed intermediate underlying graphs (containing components which consist of several boxes in all valid box decompositions).
In each test, all possible games resulting from the tested initial graph were checked, i.e., Dominator's moves were chosen according to the algorithm, and all possible legal moves were tested for each Staller move.

For efficiency reasons, the implementation differs from the algorithm described in Section \ref{section:algorithm_outline} in the following ways.
\begin{enumerate}
	\dnsitem Not all possible underlying graphs and value functions are tested, but rather a small subset which is closely related to the previous underlying graph and value function.
	\dnsitem The implementation uses a deterministic box decomposition process, rather than checking all possible box decompositions. 
	\dnsitem Dominator's move is always chosen from the vertices in the root boxes of the dense graph, and leaves are not considered except when all components of the dense graph are of size $2$.
	Note that the analysis of Dominator's moves refers only to the root boxes, and therefore there is always a move on a root box.
	\dnsitem When a semi-corrupted component is created at the end of Staller's move, the following Dominator move is chosen from the vertices of the corrupted box.
	\dnsitem If it is impossible to gain more than $6$ points on Dominator's move, and if there are several moves gaining $6$ points, ties are broken according to the following priorities:
	\begin{enumerate}
		\dnsitem Prefer to play on a white vertex of a dispensible component.
		\dnsitem Prefer a move where the resulting dense graph contains a root box of the form $BWBHH$.
		\dnsitem Prefer a move where the resulting dense graph contains a strong semi-triplet subtree or a semi-triplet subtree with a $B_3$ leaf.
		\dnsitem Prefer a move where the resulting dense graph contains a dispensible component of type $1$. 
		\dnsitem Break additional ties arbitrarily.
	\end{enumerate}
\end{enumerate}

Because we used the implementation to verify parts of the analysis as well, we did not make additional improvements, and also verified that the excess gain for Staller moves is never negative, 
and that if the excess gain of a move played by Dominator is negative, then the sum of excess gains over at most three moves starting from this move is not negative (see Theorem \ref{thm:dom_gain} for details).

The efficiency of the algorithm can be further improved using additional modifications, such as choosing the first move achieving non-negative excess gain (as described above), 
and choosing moves in a deterministic manner imitating the proofs in the analysis.

\section{Conclusions}
\label{section:conclusions}

The algorithm described for Dominator achieves the desired bound of $3 n / 5$ on all isolate-free forests, which proves Conjecture \ref{conjecture:3_5}.
The variant of the conjecture that relates to general isolate-free graphs remains open, however an upper bound of $\left\lceil 7 n / 10 \right\rceil$ is proved in \cite{kinnersley2013extremal}, and an improved bound of $2 n / 3$ is shown in \cite{bujtas2015game}. 
In \cite{bujtas2015game}, Bujt\'as further improves these results (to bounds below $3 n / 5$) for graphs with minimum degree $3$ or more.

We note that the algorithm introduced here does not perform optimally (i.e., does not achieve the game domination number) on all graphs, 
and it may be interesting to optimize the solutions and find strategies that achieve the game domination number.
Constructing a strategy for Staller may also be of interest, whether it is an optimal strategy or a strategy that performs optimally against a specific Dominator strategy.

\end{document}